\def\webDOI{http://dx.doi.org}
\def\Onera{ONERA}

\documentclass[final,3p,times]{elsarticleb}

\UseRawInputEncoding

\usepackage{stmaryrd}
\usepackage{amsmath}
\usepackage{amsfonts}
\usepackage{amssymb}
\usepackage{mathrsfs}
\usepackage{enumitem}
\usepackage{cite}
\usepackage{float}
\usepackage{soul}

\usepackage{subfigure}
\usepackage{tikz}

\usepackage{lineno}
\modulolinenumbers[5]

\usepackage[breaklinks,colorlinks=true,linkcolor=blue,citecolor=red,backref=page]{hyperref}

\journal{}

\newcommand{\Rset}{\mathbb{R}}
\newcommand{\di}{\mathrm{d}}
\newcommand{\bnabla}{{\boldsymbol\nabla}}
\newcommand{\dt}{\partial_t}
\newcommand{\TF}[1]{\widehat{#1}}
\newcommand{\norm}[1]{|#1|}
\newcommand{\normu}[1]{\left|#1\right|}
\newcommand{\dir}[1]{\hat{#1}}

\newcommand{\esp}[1]{{\mathbb E}\left\{\smash{#1}\right\}}
\newcommand{\cjg}[1]{\overline{#1}}
\newcommand{\bzero}{{\bf 0}}
\newcommand{\iexp}{\operatorname{e}}
\newcommand{\ci}{\mathrm{i}}
\newcommand{\demi}{\frac{1}{2}}

\newcommand{\rj}{r}
\newcommand{\xj}{x}
\newcommand{\yj}{y}
\newcommand{\zj}{z}

\newcommand{\rv}{{\bf\rj}}

\newcommand{\ev}{{\boldsymbol e}}

\newcommand{\vj}{v}

\newcommand{\vg}{{\bf\vj}}

\newcommand{\vref}{\vg_\iref}

\newcommand{\Mach}{M}
\newcommand{\Machv}{{\bf\Mach}}

\newcommand{\fj}{f}

\newcommand{\fv}{{\bf\fj}}

\newcommand{\iref}{0}
\newcommand{\isource}{\text{so}}

\newcommand{\roi}{\varrho}
\newcommand{\pres}{p}

\newcommand{\cel}{c}

\newcommand{\rref}{\roi_\iref}
\newcommand{\pref}{\pres_\iref}
\newcommand{\celref}{\cel_\iref}

\newcommand{\celm}{\underline{\cel}}

\newcommand{\AutoCor}{\mathcal{C}}
\newcommand{\FWd}{\Omega_d}
\newcommand{\SWd}{X_d}

\newcommand{\om}{\omega}
\newcommand{\omref}{\om_\iref}
\newcommand{\pulse}{f}
\newcommand{\Pulse}{F}

\newcommand{\Greent}{G}
\newcommand{\KGreen}{{\mathcal G}}
\newcommand{\Greenw}{\TF{\Greent}}

\newcommand{\Greenws}{\Greenw_\iref}
\newcommand{\lap}{\operatorname{\Delta}}

\newcommand{\IFunction}{\mathcal{I}}
\newcommand{\iRT}{\mathrm{RT}}
\newcommand{\iKM}{\mathrm{KM}}
\newcommand{\iCINT}{\mathrm{CINT}}
\newcommand{\iBF}{\mathrm{BF}}
\newcommand{\inclu}{\mathrm{ref}}

\newcommand{\wavel}{\lambda}
\newcommand{\wavelref}{\wavel_\iref}

\newcommand{\lcor}{\ell_c}

\newcommand{\degres}{{^{\circ}}}
\newcommand{\FWc}{\Omega_c}
\newcommand{\SWc}{X_c}
\newcommand{\Doppler}{\gamma_D}

\newcommand{\fref}[1]{Fig.~\ref{#1}}  
\newcommand{\eref}[1]{Eq.~\eqref{#1}} 
\newcommand{\sref}[1]{Sect.~\ref{#1}} 

\newcommand{\tref}[1]{Tab.~\ref{#1}}

\newcommand{\window}{w}
\newcommand{\iDAS}{\mathrm{DAS}}
\newcommand{\alert}[1]{\textcolor{black}{#1}}

\begin{document}

\begin{frontmatter}

\title{Coherent interferometric imaging in a random flow}


\author[DAAA,DIDEROT]{Etienne Gay}
\ead{etienne.gay@vo2-group.com}
\author[DAAA,CS]{Luc Bonnet}
\ead{luc.bonnet@onera.fr}
\author[DAAA]{Christophe Peyret}
\ead{christophe.peyret@onera.fr}
\author[DTIS]{\'Eric Savin\corref{corES}}
\cortext[corES]{Corresponding author. Tel.: +33(0) 146 734 645; fax: +33(0) 146 734 143}
\ead{eric.savin@onera.fr}
\author[X]{Josselin Garnier}
\ead{josselin.garnier@polytechnique.edu}
%
\address[DAAA]{\Onera/DAAA, Universit\'e Paris-Saclay, FR-92322 Ch\^atillon, France}
\address[DTIS]{\Onera/DTIS, Universit\'e Paris-Saclay, FR-91123 Palaiseau, France}
\address[DIDEROT]{LPSM, Universit\'e Paris Diderot, FR-75013 Paris, France}
\address[CS]{LMSS-Mat, CentraleSup\'elec, FR-91190 Gif-sur-Yvette, France}
\address[X]{CMAP, CNRS, \'Ecole Polytechnique, Institut Polytechnique de Paris, FR-91128 Palaiseau, France}

\begin{abstract}
This paper is concerned with the development of imaging methods to localize sources or reflectors in inhomogeneous \emph{moving} media with acoustic waves that have travelled through them. A typical example is the localization of broadband acoustic sources in a turbulent jet flow for aeroacoustic applications. The proposed algorithms are extensions of Kirchhoff migration (KM) and coherent interferometry (CINT) which have been considered for smooth and randomly inhomogeneous \emph{quiescent} media so far. They are constructed starting from the linearized Euler equations for the acoustic perturbations about a stationary ambient flow. A model problem for the propagation of acoustic waves generated by a fixed point source in an ambient flow with constant velocity is addressed. Based on this result imaging functions are proposed to modify the existing KM and CINT functions to account for the ambient flow velocity. They are subsequently tested and compared by numerical simulations in various configurations, including a synthetic turbulent jet representative of the main features encountered in actual jet flows. 
\end{abstract}


\end{frontmatter}


\section{Introduction}\label{sec:intro}

This paper is concerned with the problem of localizing acoustic sources or reflectors buried in a randomly heterogeneous moving medium.
Our main motivation is the localization of broadband acoustic sources in a turbulent jet flow for aeroacoustic applications \citep{CAN75,CAN76a,CAN76b,KRO13,SIJ14}. 
Our main goal is to introduce a wave-based imaging method that can address
such a problem.
Wave-based imaging has generally two steps. 
1) In the first step a data set is collected by an array of sensors.
In passive imaging, the goal is to localize sources and the data set consists of the signals transmitted by the sources to be localized and 
recorded by an array of receivers.
In active imaging, the goal is to localize reflectors and the data set consists of the signals transmitted by an array of controlled sources  through the medium to be imaged and recorded by an array of receivers.
2)
In the second step, the data set is processed to form an image of the medium
that is intended to show the positions of the targets (sources or reflectors).

Wave-based imaging has many modalities and applications.
In seismic imaging the goal is to estimate the locations of one or more underground sources or reflectors with a passive network of receivers (seismometers) on the surface of the earth, or the locations of reflective structures from the seismic signals generated by active sources (explosions or air guns) in reflection seismology \citep{biondi,GAR16}.
The same principle is used in medical ultrasound echography \citep{hill}, in underwater acoustics (sonar) \citep{jensen11} or in radar \citep{cheney} for instance.

Imaging sources or reflectors, when performed in a smooth medium, is effective with Kirchhoff migration (KM) \citep{BLE01}, when the background propagation velocity is known or can be estimated. The KM imaging technique consists in travel-time migration of the recorded signals, a simplified version of the reverse-time imaging technique where the time-reversed recorded signals are numerically back-propagated by solving a wave equation in a fictitious environment. The waves travel back to their origins (original sources or reflectors) and one can then form an image of them.
KM is known to be robust with respect to additive noise, such as measurement noise \citep{AMM12}. This is illustrated for example in \citep{BOR11c}. However, KM fails to image in a randomly heterogeneous (cluttered) stationary medium \citep{BOR11b}.
Indeed, when the medium is cluttered (\emph{i.e.}, when it contains unknown small-scale inhomogeneities), which is generally the case in real-life imaging, travel-time or KM does not work well, because this method in which sources transmit pulses are dependent on clear arrival times. When the medium is cluttered the recorded temporal traces are randomly shifted by the fluctuations of the speed of propagation and they contain long and noisy coda resulting from multiple scattering. The images obtained are then very noisy. Worse they are dependent on the medium, and for two different realizations of the random medium with the same statistical properties we can get two significantly different images.

Contrarily to measurement noise which induces additive independent perturbations in the recorded signals,
the noise generated by cluttered medium in the recorded signals has a complex and correlated structure, which itself generates complex and strong noise in the images formed by KM.
For these reasons it is necessary to use imaging methods that are robust with respect to the fluctuations of the medium when it is randomly heterogeneous. 
In several studies Borcea and her co-workers \citep{BOR03, BOR05, 
BOR06b, BOR06c, BOR07, 
BOR11a, BOR11b, BOR11c, BOR17} have introduced a new imaging modality called Coherent INTerferometry (CINT) whose principle is to back-propagate the cross correlations of the recorded signals, and not the signals themselves.
In \citep{BOR11b} the authors analyze the KM and CINT methods for passive and active arrays of sensors, and quantify their resolution limits and their signal-to-noise ratio (SNR).
The resolution and statistical fluctuations of the images are analyzed when the ambient environment is random and wave propagation can be modeled mainly by the distortion of the wavefront. KM loses statistical stability at an exponential rate with the propagation distance and leads to unreliable images that change unpredictably with the detailed features of the clutter. Borcea \emph{et al.} \citep{BOR03,BOR05} as well as Chan \emph{et al.} \citep{CHA99} show that CINT imaging statistically stabilizes the images. That is, images have small variance with respect to the random fluctuations of the propagation medium. 
However, the statistical stability of the CINT method comes at the detriment of some blurring of the images. The trade-off between the resolution and the stability of the CINT imaging function can be explicitly quantified \citep{BOR11b}. The results obtained so far are very encouraging and this is the reason that motivates us to extend this kind of imaging techniques to heterogeneous flows, \emph{i.e.} random moving media.

More recently, correlation-based imaging have been used to observe fast moving objects  \citep{BOR17,FOU17}. In these papers, the authors introduce a Doppler compensation parameter which is used in the imaging function to provide the necessary correction of the movement of the object to be observed. In our paper we will consider the same compensation mechanism to extend the CINT imaging algorithm to moving heterogeneous media such as jet flows. One of the most important results is that a sparse configuration of a small number of uniformly distributed receivers over an area big enough is sufficient to create an image with the same resolution as the one obtained with a dense set of receivers. It is therefore possible, with a Doppler correction, to obtain an image of a fast moving object. This opens very interesting perspectives for the objective of imaging through a random flow, which we  do in our paper. 

The rest of the paper is organized as follows. In \sref{sec:func} a model problem of acoustic wave propagation in a flow with constant velocity is addressed, starting from the linearized Euler equations with a fixed point source \alert{being either a monopole or a dipole}. Imaging functions for moving cluttered media extending the existing KM and CINT functions for quiescent cluttered media are then proposed. They introduce a phase compensation factor accounting for the Doppler shift induced by the background flow velocity. \alert{Here we also connect KM with delay-and-sum (DAS) beamforming (see \citep{BOR11a} and references therein) which is often used in aeroacoustics \citep{GRA95}}. The KM and CINT functions for quiescent media are reviewed in \sref{sec:Num:res} for configurations already considered in the literature in order to validate our numerical implementation by comparisons with the existing results in \citep{BOR05,BOR06b}. Then the proposed imaging functions are tested in \sref{sec:moving-random-medium} for moving cluttered media with a random speed of sound and either a constant or a randomly perturbed velocity of the background flow. They are applied to a numerically generated synthetic jet in \sref{sec:Jet}, where a compensation scheme for refraction of the acoustic waves at the jet sheared layers is introduced in addition to the Doppler shift compensation. A summary of this research and conclusions are finally drawn in \sref{sec:CL}.  

\section{Imaging functions in moving random media}\label{sec:func}

\subsection{Model problem}

We start by considering the problem of computing 
 the pressure field emitted by a point source in an ambient flow with stationary velocity because it has relevance to the imaging algorithms considered in the subsequent developments. The Euler equations for mass, momentum, and energy conservation in the compressible flow of an ideal fluid without friction, heat conduction or heat production read\string:
\begin{equation}\label{eq:Euler}
\begin{split}
\frac{\di\roi}{\di t} &=-\roi\bnabla\cdot\vg + m \,, \\
\frac{\di\vg}{\di t} &=-\frac{1}{\roi}\bnabla\pres +\alert{\fv} \,,\\
\frac{\di s}{\di t} &=0\,,
\end{split}
\end{equation}
where $\smash{\frac{\di}{\di t}}=\smash{\dt+\vg\cdot\bnabla}$ is the convective derivative with $\bnabla =\smash{(\frac{\partial}{\partial \xj}, \frac{\partial}{\partial \yj}, \frac{\partial}{\partial \zj})}$, $\vg(\rv,t)$ is the fluid velocity, $\roi(\rv,t)$ is the density, $\pres(\rv,t)$ is the (static) fluid pressure, $s(\rv,t)$ is the specific entropy, $m(\rv,t)$ is a specific mass source per unit time, \alert{and $\fv(\rv,t)$ is a body force per unit mass (neglecting gravity)}, at the position $\rv= (\xj,\yj,\zj)$ in $\Rset^3$ and time $t$ in $\Rset$. The flow is isentropic (\emph{i.e.} each fluid particle has constant entropy), and by the equation of state \alert{$\pres=\pres(\roi,s)$} we have\string:
\begin{equation}\label{eq:state}
\frac{\di\pres}{\di t}=\cel^2\frac{\di\roi}{\di t}\,,\alert{\quad\cel^2(\roi,s)=\left.\frac{\partial\pres}{\partial\roi}\right|_s}\,,
\end{equation}
where $\cel$ is the adiabatic speed of sound. For a perfect gas of heat capacity ratio $\gamma$, the equation of state $\smash{\pres\roi^{-\gamma}}=C$ along particle paths, where the constant $C$ may differ for each particle in isentropic flows, yields $\smash{\cel^2}=\gamma\pres/\roi$. The foregoing Euler equations are linearized about an unperturbed, stationary flow for which the pressure $\smash{\pref(\rv)}$, velocity $\smash{\vref(\rv)}$, and density $\smash{\rref(\rv)}$ do not depend on time. They satisfy\string:
\begin{equation}\label{eq:stat_flow}
\begin{split}
(\vref\cdot\bnabla)\rref &=-\rref\bnabla\cdot\vref\,, \\
(\vref\cdot\bnabla)\vref &=-\frac{1}{\rref}\bnabla\pref\,,\\
(\vref\cdot\bnabla)\pref &=\celref^2(\vref\cdot\bnabla)\rref\,,
\end{split}
\end{equation}
where $\smash{\celref(\rv)}$ stands for the sound velocity not influenced by the acoustic waves propagating in the actual flow. Linearization consists in considering that the latter is a perturbation $\smash{(\pres',\vg',\roi')}$ of the stationary flow generated by the specific mass $m$ injected to the fluid\string:
\begin{displaymath}
\begin{split}
\pres(\rv,t) &=\pref(\rv)+\pres'(\rv,t)\,, \\
\vg(\rv,t) &=\vref(\rv)+\vg'(\rv,t)\,,\\
\roi(\rv,t) &=\rref(\rv)+\roi'(\rv,t)\,. \\
\end{split}
\end{displaymath}
The primed quantities $\pres'$, $\vg'$, and $\roi'$ are the acoustic pressure, velocity, and density, respectively, of which non-linear contributions to the Euler equations \eqref{eq:Euler} are assumed negligible. In other words, these quantities are first-order increments of the zero-th order stationary quantities. They satisfy the following linearized Euler equations (LEE)\string:
\begin{equation}\label{eq:LEE}
\begin{split}
\frac{\di\roi'}{\di t}+\bnabla\cdot(\rref\vg')+\roi'\bnabla\cdot\vref &= m \,, \\
\frac{\di\vg'}{\di t} + (\vg'\cdot\bnabla)\vref + \frac{1}{\rref}\bnabla\pres' - \frac{\roi'}{\rref^2}\bnabla\pref & = \alert{\fv}\,, \\
\end{split}
\end{equation}
together with the barotropic assumption $\smash{\pres'=\celref^2\roi'}$. Here and from now on the convective derivative is defined as $\smash{\frac{\di}{\di t}}=\smash{\dt+\vref\cdot\bnabla}$, \emph{i.e.} the convective derivative within the stationary, ambient flow.

If the ambient quantities are non vanishing constant $\smash{(\pref,\vref,\rref)}$, \alert{that is, the ambient flow is uniform} which is consistent with \eref{eq:stat_flow}, we obtain by taking the convective derivative $\smash{\frac{\di}{\di t}}$ of the first equation in (\ref{eq:LEE}) and the divergence $\bnabla\cdot$ of the second equation in (\ref{eq:LEE})\string:
\begin{equation}\label{eq:convected-wave-equation}
\frac{1}{\celref^2}\frac{\di^2\pres}{\di t^2}-\lap\pres=\frac{\di m}{\di t} \alert{-\rref\bnabla\cdot\fv}\,.
\end{equation}
Here we have dropped the primes $(\cdot)'$ for convenience. \alert{The convected wave equation}\string:
\begin{displaymath} 
\frac{\di}{\di t}\left(\frac{1}{\celref^2}\frac{\di\pres}{\di t}\right)-\rref\bnabla\cdot\left(\frac{1}{\rref}\bnabla\pres\right)=\frac{\di m}{\di t} \alert{-\rref\bnabla\cdot\fv}
\end{displaymath}
first proposed by Blokhintzev \citep{BLO46a} for steady irrotational ambient flows, \alert{reduces to \eref{eq:convected-wave-equation} for a uniform flow}.
\alert{A solution can be constructed explicitly by introducing a local frame moving at the constant velocity $\smash{\vref}$, in which case \eref{eq:convected-wave-equation} reads as a classical acoustic wave equation with a vanishing ambient flow velocity. The analysis is outlined in \emph{e.g.} \citep[Chapter 5]{OST16}. First, considering a monopole source $m(\rv,t)=A(t)\smash{\delta(\rv-\rv_s)}$ of amplitude $A$ and centered at the position $\smash{\rv_s}$ and no dipole (momentum) source, the solution of \eref{eq:convected-wave-equation} reads\string:
\begin{equation}\label{eq:pres-in-flow}
\pres(\rv,t)=\frac{A'\left(t-\Doppler(\rv,\rv_s,\Machv)\frac{\normu{\rv-\rv_s}}{\celref}\right)}{4\pi\normu{\rv-\rv_s}\sqrt{1-\Mach^2+\left(\Machv\cdot\frac{\rv-\rv_s}{\normu{\rv-\rv_s}}\right)^2}}\,,
\end{equation}
where $\Machv:=\smash{\frac{\vref}{\celref}}$ is the Mach number (a vector in the present case) of the ambient flow, $\Mach=\normu{\Machv}$, and\string:
\begin{equation}\label{eq:compensation}
\Doppler(\rv,\rv',\Machv)= \frac{1}{1-\Mach^2}\left[\sqrt{1-\Mach^2+\left(\Machv\cdot\frac{\rv-\rv'}{\normu{\rv-\rv'}}\right)^2}-\Machv\cdot\frac{\rv-\rv'}{\normu{\rv-\rv'}}\right]
\end{equation}
is a Doppler compensation factor which compensates for the shift of the arrival time induced by the flow when the ambient medium moves at the constant velocity $\smash{\vref}$. We note that this factor depends on the direction of $\rv-\rv'$ but not its length. We will resort to this expression to form imaging functions in moving media in the next sections.}

\alert{Second, considering a dipole source $d(\rv,t)=A(t)\left[\smash{\delta(\rv-\rv_+)}-Ó\smash{\delta(\rv-\rv_-)}\right]$ such that $-\rref\bnabla\cdot\fv\equiv d$ with $\rv_- -\rv_+=\epsilon\ev$, $\epsilon$ the small distance between $\smash{\rv_-}$ and $\smash{\rv_+}$, $\ev$ the direction of the dipole, and no monopole (mass) source, the solution of \eref{eq:convected-wave-equation} for $\epsilon\ll\smash{\normu{\rv-\rv_+}}$ reads\string:
\begin{equation}\label{eq:pres-dipole}
\pres(\rv,t)\simeq\epsilon\ev\cdot\bnabla\left[\frac{A\left(t-\Doppler(\rv,\rv_+,\Machv)\frac{\normu{\rv-\rv_+}}{\celref}\right)}{4\pi\normu{\rv-\rv_+}\sqrt{1-\Mach^2+\left(\Machv\cdot\frac{\rv-\rv_+}{\normu{\rv-\rv_+}}\right)^2}}\right]\,.
\end{equation}
Thus the same compensation factor $\smash{\Doppler}$ for the phase arises in these expressions for either a monopole source or a dipole source}.

\subsection{Reverse-time and Kirchhoff migration}\label{subsec:Kirchhoff}

We first consider the classical reverse-time (RT) \citep{TAR84} and Kirchhoff migration (KM) \citep{BLE01} algorithms to detect and localize sources or reflectors in a homogeneous or weakly  heterogeneous medium. We adapt them to the situation of interest for us, namely the localization of such sources or reflectors in a random flow. In RT migration \citep{TAR84}, $N$ fixed sensors located at $\rv_r$, $1\leq r\leq N$, are used to localize a source (passive imaging case) or a reflector (active imaging case) at some unknown location $\smash{\rv^\star}$ in the actual (unknown) random flow by back-propagating in a fictitious flow of known characteristics the pressure fields recorded by the sensors.

\subsubsection{Passive imaging}

\begin{figure}[ht!]
\centering
\subfigure[Numerical setup]{\includegraphics[scale = 0.9]{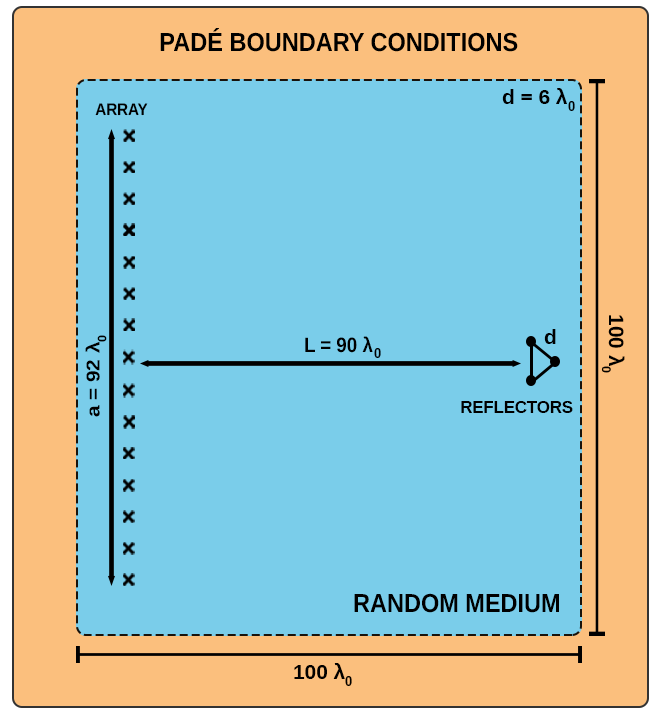}}
\subfigure[Sample of the speed of sound]{\vspace*{+10pt}\includegraphics[scale = 0.41]{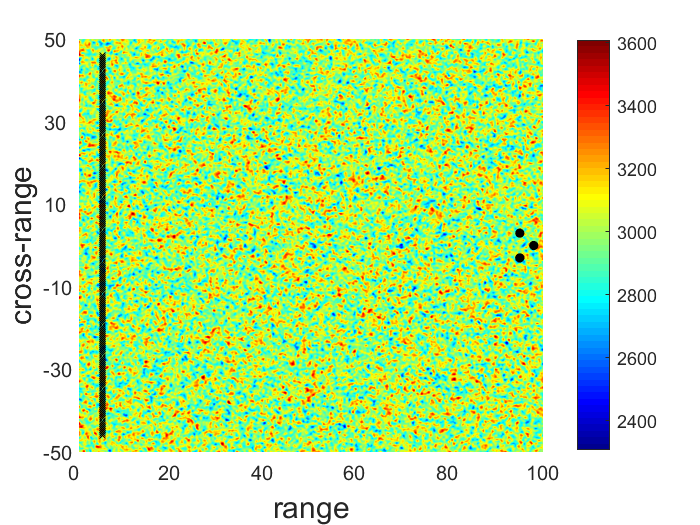}}
\caption{Typical imaging configuration in a quiescent heterogeneous medium with random speed of sound, after \citep{BOR05}. Dimensions are given in units of the central wavelength $\wavelref$ of the sources. (a) The locations of the sources (passive imaging) or reflectors (active imaging) to be imaged are shown by dots $\bullet$ and the locations of the transducers are shown by crosses $\times$. In active imaging the central transducer is used as a source. (b) Typical realization of the random speed of sound $\celref(\rv)$ with average $\celm=3000$ m/s.}\label{fg:ImgConfig}
\end{figure}

Let us first consider the case of a quiescent homogeneous medium that is a flow with $\smash{\vref}=\bzero$ and a constant speed of sound $\celref$. \alert{The Fourier transform and its inverse are defined by}\string:
\begin{equation}
\TF{\pres}(\rv,\om)=\int_\Rset \iexp^{\ci\om t} \pres(\rv,t)\,\di t\,,\quad\pres(\rv,t)=\frac{1}{2\pi}\int_\Rset \iexp^{-\ci\om t}\TF{\pres}(\rv,\om)\,\di\om\,.
\end{equation}
The time-harmonic Green's function $\Greenws(\rv,\rv',\om)$ solves the wave equation \eqref{eq:convected-wave-equation} for $\smash{\vref}=\bzero$ in the Fourier domain--the Helmholtz equation with point source at $\rv'$: 
\begin{equation}\label{eq:Helmholtz}
\frac{\om^2}{\celref^2}\Greenws+\lap\Greenws=-\delta(\rv-\rv')\,,
\end{equation}
together with the Sommerfeld radiation condition at infinity $\norm{\rv}\rightarrow+\infty$\string:
\begin{equation}\label{eq:CRS}
\left(\dir{\rv}\cdot\bnabla-\ci\frac{\om}{\celref}\right)\Greenws(\rv,\rv',\om)=\mathrm{O}\left(\frac{1}{\norm{\rv}^2}\right)\,,
\end{equation}
where $\dir{\rv}:=\smash{\frac{\rv}{\norm{\rv}}}$\string; that is, $\smash{\Greenws(\rv,\rv',\om)}\propto\smash{\iexp^{\ci\frac{\om}{\celref}\norm{\rv}}}$ at infinity. It satisfies the reciprocity property $\Greenws(\rv,\rv',\om)=\Greenws(\rv',\rv,\om)$ everywhere, and since the ambient medium has a constant speed of sound\string:
\begin{equation}\label{eq:Green0}
\Greenws(\rv,\rv',\om)=\TF{\KGreen}_\iref(\norm{\rv-\rv'},\om)=\frac{\iexp^{\ci\frac{\om}{\celref}\norm{\rv-\rv'}}}{4\pi\norm{\rv-\rv'}}\,.
\end{equation}
Passive imaging is aimed at localizing a source in a quiescent heterogeneous medium of which the actual speed of sound $\celref(\rv)$ is instead variable and imperfectly known. One starts by recording at the $N$ sensors the pressure fields emitted by the source $\Pulse(\rv,t)=\pulse(t)\smash{\delta(\rv-\rv^\star)}$. This yields the dataset $\{\pres(\rv_r,t);\,1\leq r\leq N\}$ (or its Fourier transforms $\{\TF{\pres}(\rv_r,\om);\,1\leq r\leq N\}$) constituted by the pressure fields recorded by the sensors located at $\smash{\rv_r}$, $\smash{1\leq r\leq N}$. They are subsequently time reversed, which amounts of taking the conjugates of their Fourier transforms, and back propagated in a fictitious domain of which Green's function is known, namely $\smash{\TF{\KGreen}_\iref(d,\om)}$. The RT imaging function $\IFunction_\iRT(\rv^S)$ is ultimately formed by stacking all the data\string:
\begin{equation}\label{eq:prt}
\IFunction_\iRT(\rv^S)=\frac{1}{2\pi}\sum_{r=1}^N\int_\Rset\TF{\KGreen}_\iref(\norm{\rv^S-\rv_r},\om)\cjg{\TF{\pres}(\rv_r,\om)}\,\di \om\,.
\end{equation}
Because the amplitude factor plays no role when one tries to localize sources (or reflectors), the KM function \citep{BLE01} for passive imaging, considering $\smash{\TF{\KGreen}_\iref(d,\om)}\propto\smash{\iexp^{\ci\frac{\om d}{\celref}}}$ and replacing the Green's function by this phase term in \eref{eq:prt} above, reads\string:
\begin{equation}\label{eq:Kirchhoff-source}
\begin{split}
\IFunction_\iKM(\rv^S) &=\frac{1}{2\pi}\sum_{r=1}^N\int_\Rset\iexp^{\ci\frac{\om}{\celref}\norm{\rv^S-\rv_r}}\,\cjg{\TF{\pres}(\rv_r,\om)}\,\di\om \\
&=\sum_{r=1}^N\pres\left(\rv_r,\frac{\norm{\rv^S-\rv_r}}{\celref}\right)\,,
\end{split}
\end{equation}
so that the source location can be estimated by\string:
\begin{displaymath}
\hat\rv_\isource=\arg\max_{\rv^S\in S}\IFunction_\iKM(\rv^S)\quad\text{in some search region $S$}\,.
\end{displaymath}
The range resolution of KM is $ {\celref}/{B}$, where $B$ is the source frequency bandwidth. \alert{It corresponds to the minimum distance between two sources (or reflectors in the active imaging setting) the image can identify in the range direction. The cross-range resolution is $\smash{r_c}= {\wavelref L}/{a}$ (Rayleigh resolution formula for the minimum distance between two sources or reflectors in the cross-range direction)}, where $a$ is the aperture of the sensor array, $\smash{\wavelref}$ is the source central wavelength, and $L$ is the distance from the source  to the array (range); see \fref{fg:ImgConfig} for the typical configuration of imaging sources or reflectors in a two-dimensional quiescent heterogeneous medium with random speed of sound considered in \citep{BOR05}. KM is known to work poorly when the medium is scattering, because the echoes of the sources or reflectors recorded at the array have random time shifts and a lot of delay spread--the coda--induced by the scatterers. However it is very robust with respect to additive measurement noise. 

A technique very similar to KM is delay-and-sum (DAS) beamforming, which is commonly used in aeroacoustics. DAS beamforming forms an image by \emph{summing} the signals received at the array and weighting them by weights depending on the search point $\rv^S\in S$, and \emph{delaying} the arrivals times from the search region $S$ for all receivers. That is, let $\smash{\window_r}$ be a such a real-valued weighting function centered about the receiver $\rv_r$, the DAS beamforming function for passive imaging reads:
\begin{equation}\label{eq:DAS:passive}
\IFunction_\iDAS(\rv^S)=\sum_{r=1}^N \window_r(\rv^S)\pres\left(\rv_r,\frac{\norm{\rv^S-\rv_r}}{\celref}\right)\,.
\end{equation}
For example, the weight function $\smash{\window_r}(\rv^S)=4\pi\norm{\rv^S-\rv_r}$ compensates for the amplitude decay of the signals which has been ignored in the KM imaging function \eqref{eq:Kirchhoff-source}. Adaptive weights which depend on the signal statistics can also be considered \citep{GRA95}. The connections between KM, DAS beamforming and CINT, and their resolution properties are discussed in details in \citep{BOR11a}. 
In particular, KM and DAS beamforming have similar poor performances in random media.

Now in view of the results \eqref{eq:pres-in-flow} and \eqref{eq:pres-dipole}, it is proposed to modify the foregoing KM imaging function \eqref{eq:Kirchhoff-source} to include the Doppler factor $\smash{\Doppler(\rv_r,\rv^S,\Machv)}$ of \eref{eq:compensation} which compensates for the shift of the arrival time between the source and the receiver induced by the flow. The compensated KM imaging function (\ref{eq:Kirchhoff-source}) then reads\string:
\begin{equation}\label{eq:Kirchhoff-in-flow}
\IFunction_\iKM(\rv^S)=\sum_{r=1}^N\pres\left(\rv_r,\Doppler(\rv_r,\rv^S,\Machv)\frac{\norm{\rv^S-\rv_r}}{\celref}\right)\,,
\end{equation}
where one observes that $\smash{\Doppler(\rv_r,\rv^S,\bzero)}=1$ and the imaging function \eqref{eq:Kirchhoff-source} for a quiescent medium is recovered. \alert{Since the amplitudes are disregarded, this modified imaging function is aimed at localizing either monopole or dipole sources}. Note also that the same Doppler compensation factor has been considered recently in \citep{BOR17,FOU17} for imaging small fast moving objects revolving around the Earth at low orbits.

\subsubsection{Active imaging}

In active imaging, the network of sensors is used for detecting reflectors. One or several sensors $\rv_s$, $\smash{1\leq s\leq N_{\rm s}}$, transmit signals which are recorded after a round trip to the reflectors. In this context the dataset $\{\pres(\rv_r,t;\rv_s);\,1\leq r\leq N_{\rm r},\,1\leq s\leq N_{\rm s}\}$ (or its Fourier transforms $\{\smash{\TF{\pres}(\rv_r,\om;\rv_s)};\smash{\,1\leq r\leq N_{\rm r},\,1\leq s\leq N_{\rm s}\}}$) is constituted by the pressure fields recorded by the sensors located at $\smash{\rv_r}$, $\smash{1\leq r\leq N_{\rm r}}$, when the sensors located at $\smash{\rv_s}$, $\smash{1\leq s\leq N_{\rm s}}$, act as active sources one after the other one. The KM function in active imaging is written\string:
\begin{equation}\label{eq:Kirchhoff-source:actif}
\IFunction_\iKM(\rv^R) =\sum_{r=1}^{N_{\rm r}}\sum_{s=1}^{N_{\rm s}}\pres\left(\rv_r,\frac{\norm{\rv_s-\rv^R}}{\celref}+\frac{\norm{\rv^R-\rv_r}}{\celref};\rv_s\right)\,,
\end{equation}
and:
\begin{equation}
\hat\rv_\inclu
=\arg\max_{\rv^R\in S}\IFunction_\iKM(\rv^R)\quad\text{in some search region $S$}\
\end{equation}
is the estimated position of the reflector.

The DAS beamforming function for active imaging reads:
\begin{equation}\label{eq:DAS:active}
\IFunction_\iDAS(\rv^R)=\sum_{r=1}^{N_{\rm r}}\sum_{s=1}^{N_{\rm s}} \window_{rs}(\rv^R)\pres\left(\rv_r,\frac{\norm{\rv_s-\rv^R}}{\celref}+\frac{\norm{\rv^R-\rv_r}}{\celref};\rv_s\right)\,.
\end{equation}
For example, the weight function $\smash{\window_{rs}}(\rv^R)=4\pi\norm{\rv_s-\rv^R}\times 4\pi\norm{\rv^R-\rv_r}$ compensates for the amplitude decay of the signals which has been ignored in the KM imaging function \eqref{eq:Kirchhoff-source:actif}. Again, KM, DAS beamforming and CINT are compared in \citep{BOR11a}.

As for passive imaging, the foregoing \alert{KM imaging function \eqref{eq:Kirchhoff-source:actif}} is modified to include \alert{Doppler compensation factors $\smash{\Doppler(\rv_s,\rv^R,\Machv)}$ and $\smash{\Doppler(\rv^R,\rv_r,\Machv)}$} when the ambient medium moves at an average velocity $\smash{\vref}\neq\bzero$. In active imaging the KM imaging function (\ref{eq:Kirchhoff-source:actif}) now reads\string:
\begin{equation}\label{eq:Kirchhoff-in-flow:actif}
\IFunction_\iKM(\rv^R) = \sum_{r=1}^{N_{\rm r}}\sum_{s=1}^{N_{\rm s}}\pres\left(\rv_r,\Doppler(\rv_s,\rv^R,\Machv)\frac{\norm{\rv_s-\rv^R}}{\celref}+\Doppler(\rv^R,\rv_r,\Machv)\frac{\norm{\rv^R-\rv_r}}{\celref};\rv_s\right)\,.
\end{equation}
The first $\smash{\Doppler}$ compensates for the time shift induced by the flow, for the travel from the source to the reflector. The second $\smash{\Doppler}$ compensates for this shift for the travel from the reflector to the receiver.

\subsection{Coherent interferometric (CINT) imaging}\label{subsec:CINT}

Coherent interferometric imaging (CINT) \citep{BOR03, BOR05, 
BOR06b, BOR06c, BOR07, 
BOR11a, BOR11b, BOR11c} of sources or reflectors uses finite-aperture arrays alike to localize in a cluttered medium a source (passive imaging) or a reflector (active imaging).
 It consists in back propagating selected empirical correlations of the pressure fields, rather than the pressure fields themselves, in a fictitious medium in order to alleviate the statistical instability of KM with clutter. Indeed, forming the correlations enables to partially cancel the incoherent random phase shifts of the signals and thus enhance statistical stability.
We refer to \fref{fg:ImgConfig} for a typical configuration \alert{in two dimensions, with a linear array $A=\smash{[-\frac{a}{2},\frac{a}{2}]}\times\{0\}$ of $N$ sensors and a target (source or reflector) that is at the position $(0,L)$}, where $L$ is the distance (range) from the array.

\subsubsection{Passive CINT imaging}

The empirical cross-correlation of the recorded pressure fields at the sensors $\smash{\rv_q}$ and $\smash{\rv_r}$, $1\leq q,r\leq N$, reads in the Fourier domain\string:
\begin{equation}\label{eq:AutoCor}
\AutoCor_F(\rv_q,\rv_r,\om,\om')=\TF{\pres}(\rv_q,\om)\cjg{\TF{\pres}(\rv_r,\om')}\,.
\end{equation}
Then these empirical cross-correlations are back-propagated in a fictitious (\emph{e.g.} homogeneous) medium, of which Green's function is again $\smash{\TF{\KGreen}_\iref(d,\om)}\propto\smash{\iexp^{\ci\frac{\om d}{\celref}}}$. Besides, the cross-correlations are computed locally in frequency and space and not over the whole frequency range and for all pairs of sensors. The CINT function for passive imaging in a quiescent medium reads \citep{BOR05}\string:
\begin{equation}\label{eq:CINT-source}
\IFunction_\iCINT(\rv^S;\FWd,\SWd)=\sum_{\scriptsize\begin{array}{c}q,r=1\\\norm{\rv_q-\rv_r}\leq \SWd\end{array}}^N\iint_{\norm{\om-\om'}\leq\FWd}\AutoCor_F(\rv_q,\rv_r,\om,\om')\iexp^{-\ci\frac{\om}{\celref}\norm{\rv_q-\rv^S}+\ci\frac{\om'}{\celref}\norm{\rv_r-\rv^S}}\,\di\om\di\om'\,,
\end{equation}
such that the source location
 can be estimated by\string:
\begin{displaymath}
\hat\rv_\isource
=\arg\max_{\rv^S\in S}\IFunction_\iCINT(\rv^S;\FWd,\SWd)\quad\text{in some search region $S$}\,.
\end{displaymath}
The range resolution of CINT is $ {\celref}/{\FWd}$ (the usual range resolution formula with the effective bandwidth $\smash{\FWd}<B$ where $B$ is the source bandwidth), and its cross-range resolution is $ {\wavelref L}/{\SWd}$ (the Rayleigh resolution formula with an effective array diameter $\smash{\SWd}<a$) \citep{BOR05,BOR06b}. Ideally the frequency window $\smash{\FWd}$ is chosen as the decoherence frequency $\smash{\FWc}$, \emph{i.e.} the frequency gap beyond which the frequency components of the recorded pressure fields are no longer correlated. Likewise, the spatial window $\smash{\SWd}$ is ideally chosen as the decoherence length $\smash{\SWc}$, \emph{i.e.} the sensor gap beyond which the recorded pressure fields are no longer correlated. The decoherence parameters $\smash{\FWc}$ and $\smash{\SWc}$ depend on the statistical properties of the fluctuations in the random medium, the range $L$, and the central frequency. Actually these parameters are not known since the random medium is also unknown, so the CINT parameters $\smash{\FWd}$ and  $\smash{\SWd}$ have to be determined from the data by the imaging process itself. Alternatively, an optimal set of parameters may be determined by the adaptive algorithm developed in \citep{BOR06b}.

\alert{We note that the spatial cut-off induced by summing nearby receivers within the spatial window $\smash{\SWd}$ roughly corresponds to the choice of some specific weight functions $\smash{\window_r}$ in the beamforming function in reception \eqref{eq:DAS:passive}, selecting sub-apertures near the receivers. However the frequency cut-off induced by integrating nearby frequencies within the frequency window $\smash{\FWd}$ is absent in \eqref{eq:DAS:passive}. As outlined in \citep{BOR11a}, it can only be implemented in the form of a quadratic function as in the CINT function \eqref{eq:CINT-source}. It plays an important role in selecting the coherent contributions of the recorded signals, thus making CINT imaging more effective in cluttered media. Also the window parameters $\smash{\SWd}$ and $\smash{\FWd}$ are common for all pairs of receivers, whereas the weight functions in DAS beamforming have to be estimated for each search point $\rv^S$. As such, the adaptive CINT imaging procedure outlined in \citep{BOR06b} is based on the quality of the image solely rather than the statistics of the signals \citep{GRA95}}.

\alert{In view of the results \eqref{eq:pres-in-flow} and \eqref{eq:pres-dipole}}, it is proposed to modify the foregoing CINT imaging function (\ref{eq:CINT-source}) for localizing a source to include the Doppler compensation factor (\ref{eq:compensation}). The compensated CINT function for passive imaging in a moving medium then reads\string:
\begin{equation}\label{eq:CINT-in-flow}
\IFunction_\iCINT(\rv^S;\FWd,\SWd)= \sum_{\scriptsize\begin{array}{c}q,r=1\\ \norm{\rv_q-\rv_r}\leq\SWd\end{array}}^N\!\!\!\!\iint_{\norm{\om-\om'}\leq\FWd} \AutoCor_F(\rv_q,\rv_r,\om,\om') \iexp^{-\ci\frac{\om}{\celref}\Doppler(\rv_q,\rv^S,\Machv)\norm{\rv_q-\rv^S}+\ci\frac{\om'}{\celref}\Doppler(\rv_r,\rv^S,\Machv)\norm{\rv_r-\rv^S}}\,\di\om\di\om'\,.
\end{equation}
{\bf Remark}: The unfiltered, full CINT function (obtained by letting $\SWd,\FWd \to \infty$)  is\string:
\begin{align}
\nonumber
\IFunction_\iBF(\rv^S) &= \sum_{q,r=1}^{N} \iint_{\Rset^2}\AutoCor_F(\rv_q,\rv_r,\om,\om')\iexp^{-\ci\frac{\om}{\celref} \Doppler(\rv_q,\rv^S,\Machv) \norm{\rv_q-\rv^S}+\ci\frac{\om'}{\celref} \Doppler(\rv_r,\rv^S,\Machv) \norm{\rv_r-\rv^S}}\,\di\om\di\om' \\
&=(2\pi)^2 \norm{\IFunction_\iKM(\rv^S)}^2\,,
\label{eq:cintfull}
\end{align}
that is, cross-correlations of all pairs of sensors and frequencies are migrated and summed up in \eref{eq:CINT-in-flow}. Thus the same image as with KM \eqref{eq:Kirchhoff-in-flow} is formed. This shows in particular that the full migration of all cross-correlations does not work if the medium is cluttered.

\subsubsection{Active CINT imaging}

CINT imaging can be used to localize reflectors alike. Among the $N$ transducers of the array $A$, $\smash{N_{\rm s}}$ are used as sources located at $\smash{\rv_s}$, $\smash{1\leq s\leq N_{\rm s}}$, and $\smash{N_{\rm r}}$ are used as receivers (sensors) located at $\smash{\rv_r}$, $\smash{1\leq r\leq N_{\rm r}}$ (sources and receivers may be collocated). The dataset of recorded pressure fields is denoted by $\smash{\{\pres(\rv_r,t;\rv_s)};\,1\leq r\leq N_{\rm r},\,1\leq s\leq N_{\rm s}\}$. The CINT function for active imaging in a quiescent medium reads\string:
\begin{multline}\label{eq:CINT-reflector}
\IFunction_\iCINT(\rv^R;\FWd,\SWd)= \\
\sum_{\scriptsize\begin{array}{c}r,r'=1\\\norm{\rv_r-\rv_{r'}}\leq \SWd\end{array}}^{N_{\rm r}}\!\!\sum_{\scriptsize\begin{array}{c}s,s'=1\\\norm{\rv_s-\rv_{s'}}\leq \SWd\end{array}}^{N_{\rm s}} \iint_{\norm{\om-\om'}\leq\FWd}\TF{\pres}(\rv_r,\om;\rv_s)\cjg{\TF{\pres}(\rv_{r'},\om';\rv_{s'})} \iexp^{-\ci\frac{\om}{\celref}\left(\norm{\rv_r-\rv^R}+\norm{\rv^R-\rv_s}\right)+\ci\frac{\om'}{\celref}\left(\norm{\rv_{r'}-\rv^R}+\norm{\rv^R-\rv_{s'}}\right)}\,\di\om\di\om'\,,
\end{multline}
so that the reflector location
can be estimated by\string:
\begin{displaymath}
\hat\rv_\inclu
=\arg\max_{\rv^S\in S}\IFunction_\iCINT(\rv^S;\FWd,\SWd)\quad\text{in some search region $S$}\,.
\end{displaymath}

Along the same lines as in the passive imaging case, we modify the CINT function (\ref{eq:CINT-reflector}) for active imaging in a moving medium as\string:
\begin{multline}\label{eq:CINT-in-folw-reflector}
\IFunction_\iCINT(\rv^S;\FWd,\SWd)= \sum_{\scriptsize\begin{array}{c}r,r'=1\\\norm{\rv_r-\rv_{r'}}\leq \SWd\end{array}}^{N_{\rm r}}\sum_{\scriptsize\begin{array}{c}s,s'=1\\\norm{\rv_s-\rv_{s'}}\leq \SWd\end{array}}^{N_{\rm s}}\iint_{\norm{\om-\om'}\leq\FWd}\TF{\pres}(\rv_r,\om;\rv_s)\cjg{\TF{\pres}(\rv_{r'},\om';\rv_{s'})} \\
\times\iexp^{-\ci\frac{\om}{\celref}\left(\Doppler(\rv_r,\rv^S,\Machv)\norm{\rv_r-\rv^S}+\Doppler(\rv^S,\rv_s,\Machv)\norm{\rv^S-\rv_s}\right)} \iexp^{+\ci\frac{\om'}{\celref}\left(\Doppler(\rv_{r^\prime},\rv^S,\Machv)\norm{\rv_{r^\prime}-\rv^S}+\Doppler(\rv^S,\rv_{s^\prime},\Machv)\norm{\rv^S-\rv_{s^\prime}}\right)}\,\di\om\di\om'\,.
\end{multline}

\section{Imaging in a quiescent random medium}\label{sec:Num:res}

In this section the imaging algorithms outlined in \sref{sec:func} are first applied to the case of a quiescent random medium. The same two-dimensional configuration as in \citep{BOR05,BOR06b} and shown in \fref{fg:ImgConfig} is considered in order to validate the numerical solver used throughout the simulations for both quiescent and moving media. The array data $\{\pres(\rv_r,t);\,1\leq r\leq N\}$ 
or $\{\pres(\rv_r,t;\rv_s);\,1\leq r\leq N_{\rm r}, \, 1\leq s\leq N_{\rm s}\}$  of recorded pressure fields is generated numerically using the computer code SPACE \citep{DEL05,PEY17} for computational fluid dynamics (CFD) and computational aero-acoustics (CAA). It solves the Euler equations (\ref{eq:Euler}) and LEE \eqref{eq:LEE} written as first-order symmetric systems by the nodal discontinuous Galerkin (DG) finite-element method \citep{HES08},  using flux splitting techniques and an explicit Runge-Kutta time-integration scheme. In the typical configuration of \fref{fg:ImgConfig} the propagation medium is considered to be infinite in all directions, thus non-reflective, Pad\'e boundary conditions \citep{KEC05} surround the computational domain for the numerical simulations. This absorbing layer allows to limit the numerical echo induced by the boundaries of the finite-element mesh. All computations are performed using the parallel scalar cluster of \Onera. This cluster has a total of 17,360 cores for a peak performance of 667 Tflop/s. The post-processing of all results is done using Matlab$^\copyright$. The imaging functions are normalized and their range is $[0,1]$. In addition, in order to compare the results obtained by the KM and CINT algorithms, the CINT imaging function $\norm{\IFunction_\iCINT}$ and the squared KM imaging function $\norm{\IFunction_\iKM}^2$ are formed because $\norm{\IFunction_\iCINT} \propto \norm{\IFunction_\iKM}^2$ as we have seen in \eqref{eq:cintfull}. \alert{In \sref{subsec:Rand:Fix:med} below the model setup for numerical simulations is introduced, and in \sref{sec:mentionless:Acti:imag} the KM and CINT active imaging functions are tested for localizing small reflectors in this configuration. Passive imaging functions have also been tested for localizing point sources, yielding satisfactory results as well which are not presented here}.

\subsection{Model setup for numerical simulations}\label{subsec:Rand:Fix:med}

The setup for numerical validation is shown in \fref{fg:ImgConfig} where dimensions are given in terms of the central wavelength $\wavelref$. The horizontal axis is the range and the vertical axis along which the array is aligned is the cross-range. This array contains $N=185$ transducers at a distance $\wavelref/2$ from each other, with an aperture $a=92\smash{\wavelref}$. The object to be imaged is at a range $L=90\wavelref$ and zero cross-range, and it is either constituted by three sources (two at the forefront $6\wavelref$ apart from each other and one in the rear at $3\wavelref$ from the others) emitting the same signals $t\mapsto\pulse(t)$ simultaneously in the passive imaging configuration, or three non penetrable disks of radius $ {\wavelref}/{2}$ centered at the same points with homogeneous Neumann boundary conditions in the active imaging configuration. In this latter case, the central transducer of the array emits the probing pulse $t\mapsto\pulse(t)$ and all other transducers and this one are used as receivers. The pulse is (the derivative of a Gaussian):
\begin{equation}\label{eq:pulse}
\pulse(t)=-\omref^2 t \iexp^{-\demi\omref^2 t^2}\,,
\end{equation}
where $\omref=2\pi f_\iref$ is the central (angular) frequency such that $\wavelref=\smash{\celm/f_\iref}$ and $\celm$ is the average speed of sound. We choose $\smash{f_\iref}=1$ kHz 
as in \citep{BOR05}.
For this central frequency, the bandwidth of the signal is $[0.6,1.3]$ kHz. All array data in the bandwidth $[0,1.5]$ kHz have been subsequently post-processed to build the KM and CINT imaging functions.

To simulate the KM and CINT imaging processes in clutter, we consider a quiescent random medium with $\vref=\bzero$, $\rref=\text{constant}$, and a random speed of sound $\smash{\celref(\rv)}=\smash{\celm(1+\sigma\mu(\rv))}$, where $\smash{\celm}=\text{constant}$ is the average value, $\sigma$ is the relative standard deviation, and $(\mu(\rv),\,\rv\in\Rset^2)$ is a homogeneous (stationary), zero-mean second-order random process. A Gaussian covariance model is chosen for that process, namely: 
\begin{equation}\label{eq:Markov:cor}
\esp{\mu(\rv)\mu(\rv')}=\exp\left[-\demi\left(\frac{\norm{\rv-\rv'}}{\lcor}\right)^2\right]\,,
\end{equation}
corresponding to smooth samples. Here $\esp{\cdot}$ stands for the mathematical expectation, and $\smash{\lcor}$ is the correlation length of the random fluctuations $\mu$ of the speed of sound. These fluctuations are simulated by a random Fourier series of the stationary process $\mu(\rv)$; see for example \citep{POI89,SHI71}. A typical realization of the random speed of sound is shown in \fref{fg:ImgConfig} for a correlation length $\smash{\lcor}$ which is one-half the central wavelength $\wavelref$ for the sources, and a standard deviation $\sigma=3\%$. The cluttered medium in \fref{fg:ImgConfig} and subsequent simulations corresponds to an average speed of sound $\smash{\celm}=3000$ m/s and wavelength $\smash{\wavelref}=3$ m at the central frequency $\smash{f_\iref}=1$ kHz. The correlation length is $\smash{\lcor} = \smash{\wavelref/2} = 1.5$ m. This is a delicate and interesting situation to deal with, since this wavelength is \emph{a priori} required to image objects of its size and the clutter in which they are embedded has also the same characteristic length.

\begin{figure}[ht!]
\centering
\includegraphics[scale=0.35]{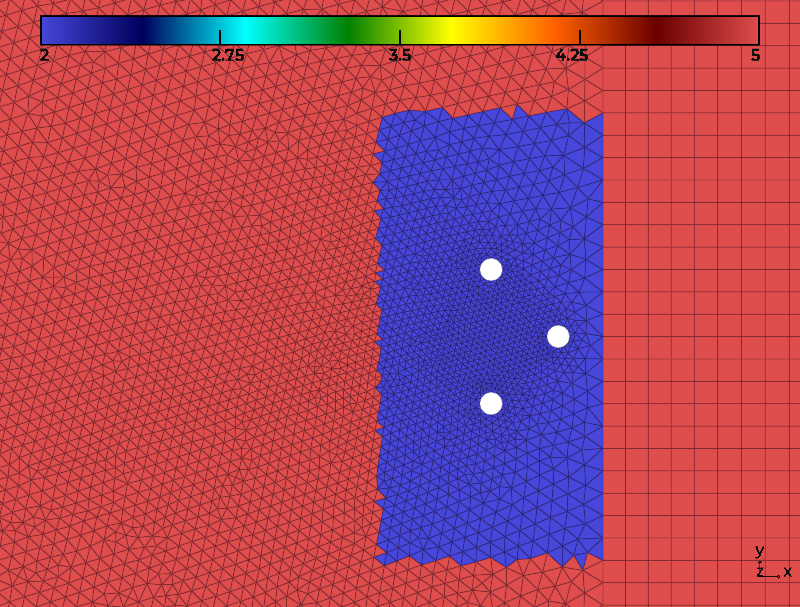}
\caption{Mesh used for the validation of the KM and CINT imaging functions in a quiescent and moving random medium. Zoom around the three reflectors to be imaged in the active imaging configuration. The blue elements have polynomial interpolation order $2$ and the red elements have polynomial interpolation order $5$.}\label{fig:maillagemedium}
\end{figure}

The same mesh is used for all our DG simulations in this section and the subsequent \sref{sec:moving-random-medium} where a moving random medium is considered. It consists in triangular and quadrangular elements with 3 and 4 vertices respectively, and includes about $1.7\times 10^5$ vertices and $1.8\times 10^5$ elements. \fref{fig:maillagemedium} shows a zoom on the mesh around the three reflectors to be imaged in the active imaging configuration. For the central frequency $\smash{f_\iref} = 1000$ Hz and the average speed of sound $\smash{\celm} = 3000$ m/s, this corresponds to about 5 to 10 nodes per wavelength in the most refined area.
We use a mixed finite element mesh with refined elements of polynomial interpolation order $2$ close to the reflectors (shown in blue on \fref{fig:maillagemedium}) and coarser elements of polynomial interpolation order $5$ far from the reflectors (shown in red on \fref{fig:maillagemedium}) to evade possible numerical instabilities for a moving random medium. In \citep{BOR05}, a piecewise polynomial interpolation of degree $1$ on regular four-node quadrilateral elements with 30 nodes per wavelength has been considered, as well as a mixed finite element method for spatial integration and perfectly matched layers (PML) \citep{BER94} to surround the computational domain. They also used a second-order leapfrog scheme \citep{BEC97,BEC00} for temporal integration, while we use a second-order Runge-Kutta explicit scheme \citep{DEL05,PEY17}. Some differences will thus be observed in the resolution of the results obtained here compared to those obtained in \citep{BOR05}. However, the DG method considered in this research for the simulations allows to increase the order of the mesh and polynomial interpolation straightforwardly. 

\begin{figure}[ht!]
\begin{center}
\subfigure[Constant speed of sound]{\includegraphics[scale = 0.45]{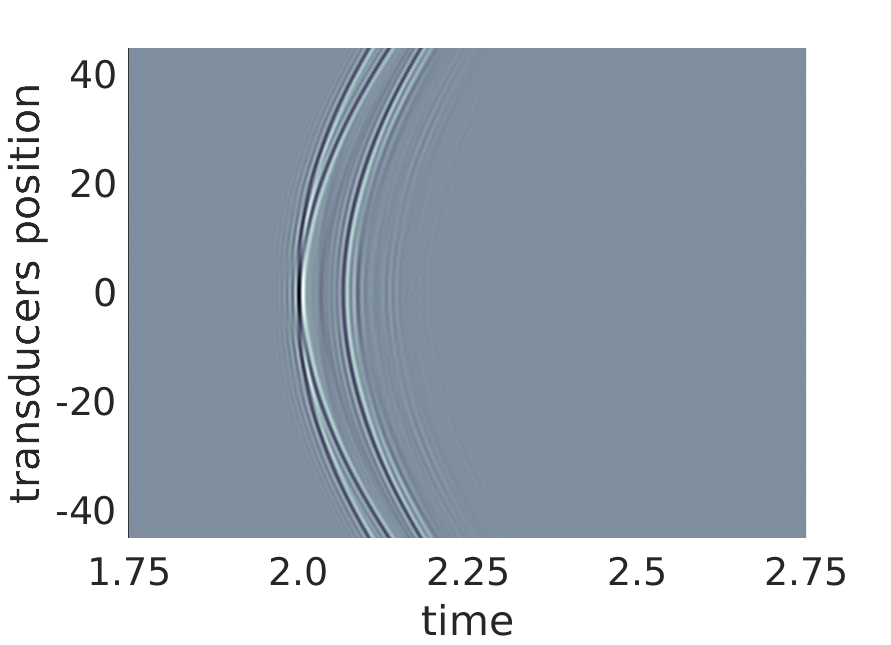}}
\subfigure[Random speed of sound]{\includegraphics[scale = 0.45]{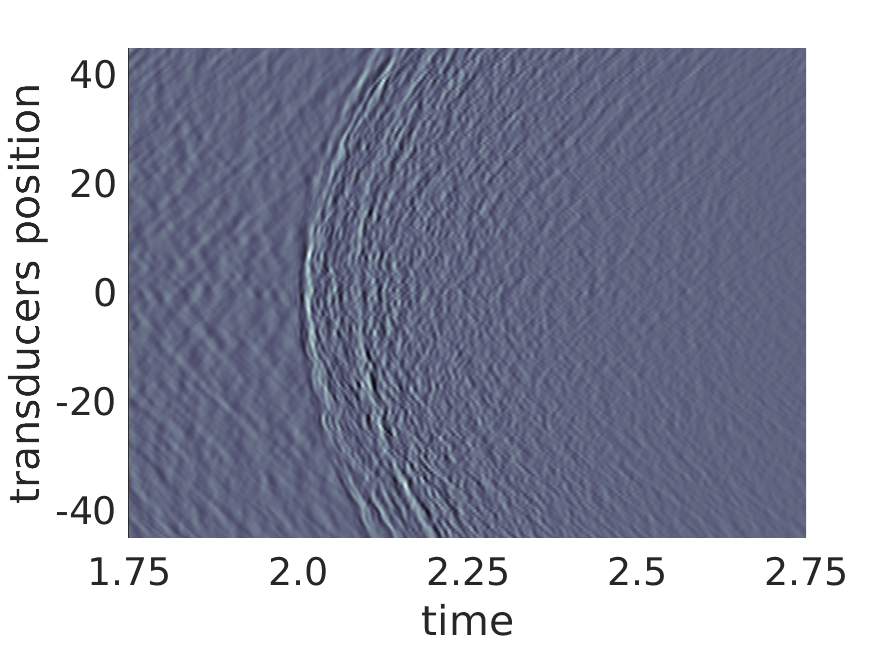}}
\end{center}
\caption{Active imaging in a quiescent medium in the configuration of \fref{fg:ImgConfig}. (a) Time traces recorded at the receivers for a constant speed of sound $\celm$ (each line plots the signal recorded by one receiver). (b) Time traces recorded at the receivers for a random speed of sound with average $\celm$ and standard deviation $\sigma=3\%$. The horizontal axis is time (\alert{in units of $L/\smash{\celm}$}) and the vertical axis is the array transducer location (\alert{in units of $\wavelref$}).}\label{fg:sig-nofh}
\end{figure}

Numerically generated data recorded at the array are shown in \fref{fg:sig-nofh} for the active imaging configuration (localization of reflectors). The central sensor of the array is used as a source, and all the sensors of the array are used as as receivers. In the active imaging configuration wavefronts have travelled twice the distance from the array to the reflectors before being recorded. We clearly distinguish the echoes from the three reflectors in the homogeneous case (\fref{fg:sig-nofh}, left). In addition, the signals are symmetric with respect to the central receiver. The effects of the heterogeneity of the speed of sound in the random medium are clearly visible on these time traces (\fref{fg:sig-nofh}, right). It is much more difficult to distinguish the wavefronts and the echoes from the three reflectors since the signals are noisy. The causes of this noise are the multiple scattering of acoustic waves on the inhomogeneities present in the random medium. Even if the typical amplitude of the inhomogeneities is weak ($\sigma\ll 1$), the range $L$ is large with respect to $\wavelref$ and $\lcor$ and there are significant multiple scattering effects. In addition, the noise will be different from one realization of the medium to another. 

\subsection{Active imaging}\label{sec:mentionless:Acti:imag}

\begin{figure}[ht!]
\centering
\includegraphics[scale=0.45]{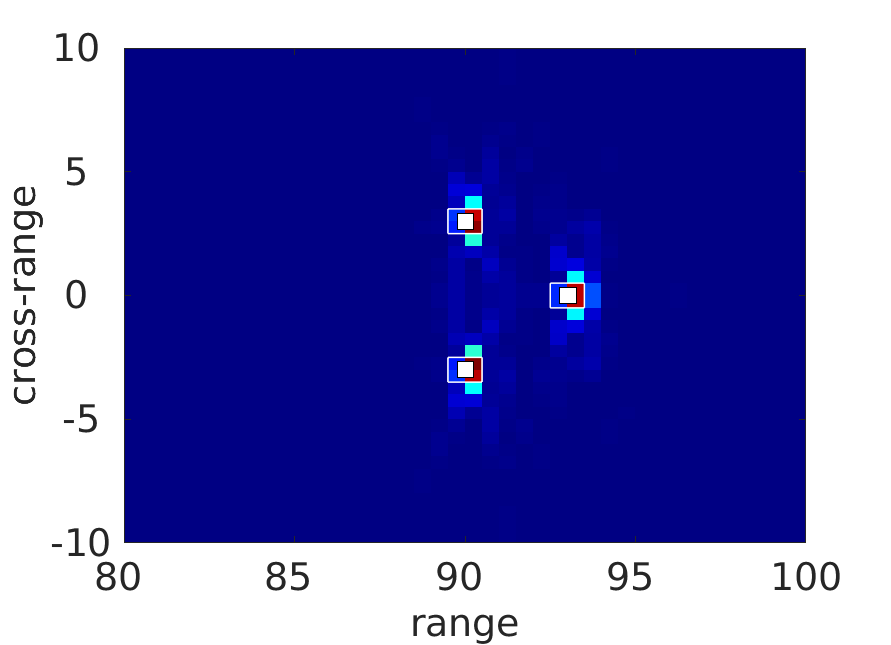}
\caption{KM active imaging in a quiescent medium in the configuration of \fref{fg:ImgConfig} with a constant speed of sound $\celm$. The reflectors to be localized are shown by white squares. Range and cross-range dimensions are in units of $\wavelref$.}\label{KIRC-nofh}
\end{figure}

We first consider the localization of three reflectors using the KM imaging function \eqref{eq:Kirchhoff-source:actif} and the CINT imaging function \eqref{eq:CINT-reflector} with $\smash{N_{\rm r}}=N=185$ and $\smash{N_{\rm s}}=1$. The search domain is a square of size $\smash{20\wavelref\times20\wavelref}$ centered around $(90\wavelref,0)$. The results obtained for the (square) KM image in a homogeneous medium with constant speed of sound are shown in \fref{KIRC-nofh} for a pixel size of $\smash{\wavelref/2}$. \alert{The theoretical range resolution is $\smash{2f_\iref/B}\simeq 1$ in pixel units, and the theoretical cross-range resolution is $2L/a\simeq 2$}. The three reflectors are actually found in the middle of the search region and their positions are precisely localized by the brightest pixels of the KM imaging function. This result validates both the efficiency of KM active imaging in homogeneous media, and the ability of our code to correctly simulate acoustic wave propagation with negligible numerical dispersion.

\begin{figure}[ht!]
\begin{center}
\subfigure[KM image in random medium \#1]{\includegraphics[scale = 0.45]{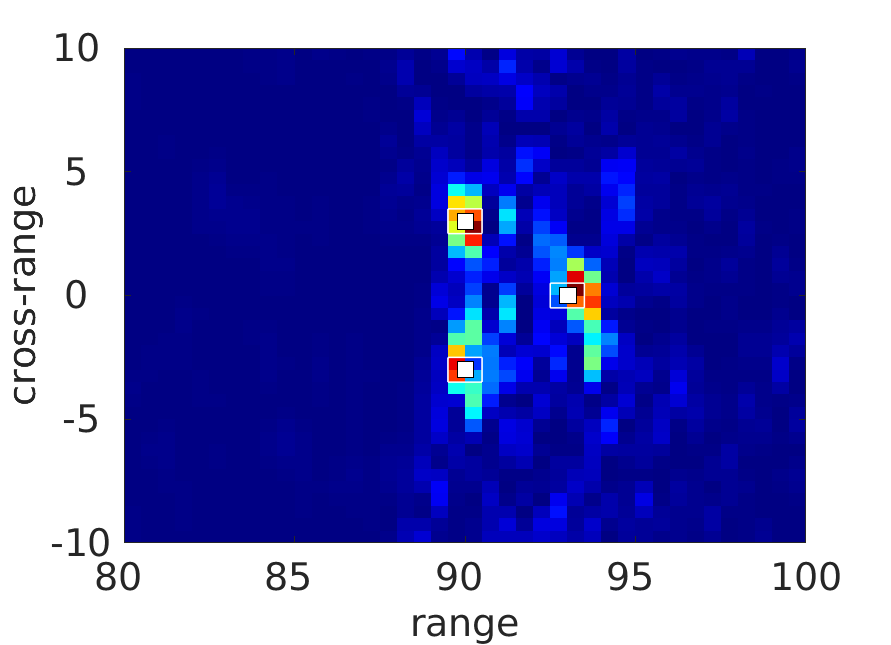}}
\subfigure[KM image in random medium \#2]{\includegraphics[scale = 0.45]{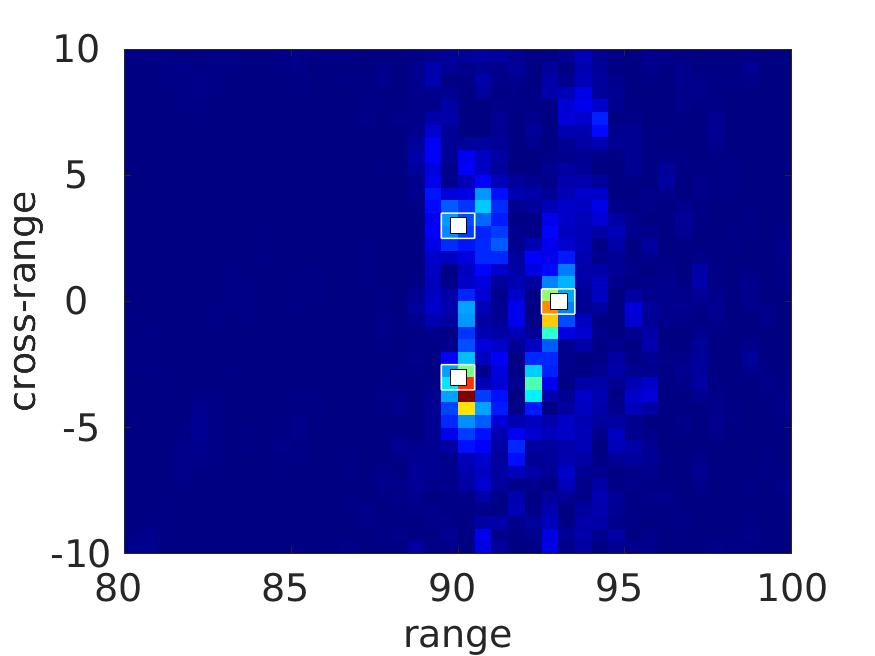}}\\
\subfigure[CINT image in random medium \#1]{\includegraphics[scale = 0.45]{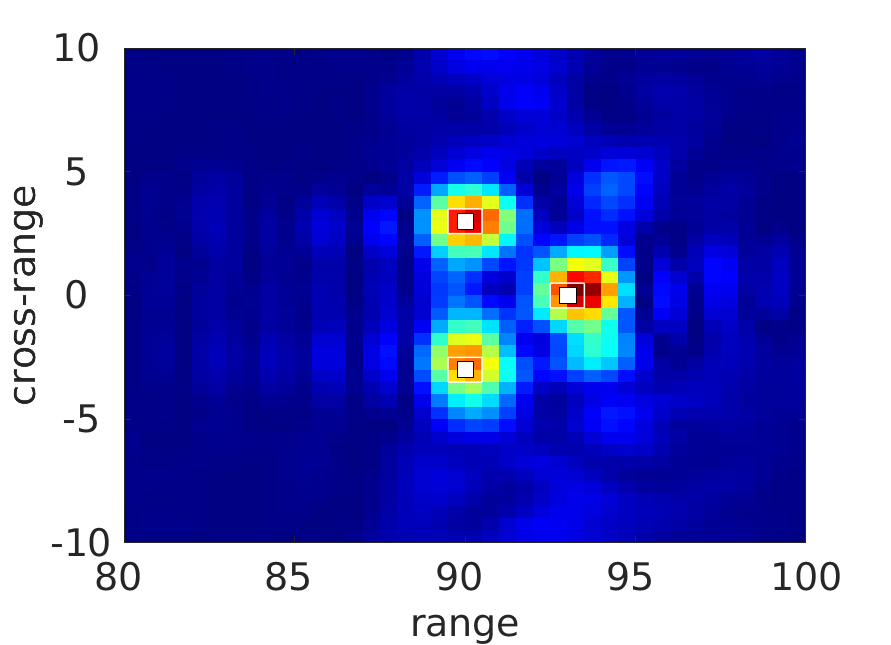}}
\subfigure[CINT image in random medium \#2]{\includegraphics[scale = 0.45]{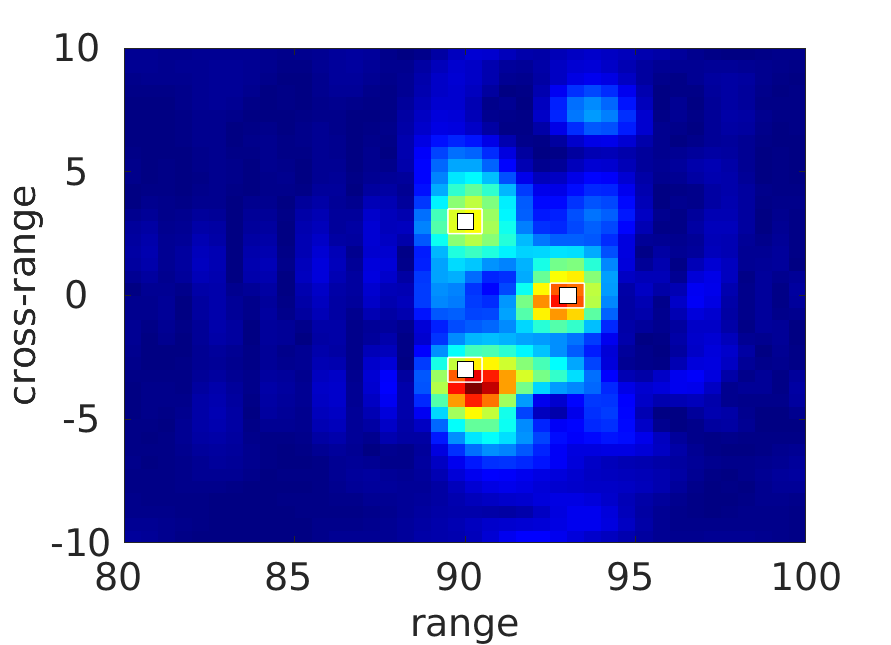}}
\end{center}
\caption{Active imaging in a quiescent medium in the configuration of \fref{fg:ImgConfig} with a random speed of sound of average $\celm$ and standard deviation $\sigma = 3 \%$. Comparison of KM and CINT images for two realizations of the random medium: (a) Squared KM image for the first realization of the medium; (b) Squared KM image for the second realization of the medium; (c) CINT image for the first realization of the medium; (d) CINT image for the second realization of the medium. The CINT parameters $\smash{\FWd} = 0.1B$ and $\smash{\SWd}= 0.8a$ are used for both media. The reflectors to be localized are shown by white squares {\tiny$\square$}. Range and cross-range dimensions are in units of $\wavelref$.}\label{fig:resnoflow}
\end{figure}

The results obtained for the (square) KM images and CINT images in a random medium are shown in \fref{fig:resnoflow} for two different realizations of the random speed of sound with the same average and standard deviation. The ineffectiveness of the KM imaging algorithm is apparent in that we observe noisy images that are different from one realization of the medium to another, and are thus statistically unstable. Indeed the KM algorithm back-propagates all the noise contained in the signals recorded by the receivers in a fictitious, homogeneous ambient medium. 
We do not get sharp peaks around the positions of the three reflectors as in the case of a homogeneous medium, because the random phases present in the signals in the Fourier domain are not satisfactorily compensated by back-propagation (i.e. by the phases of the homogeneous Green's functions), contrarily to what is done when empirical cross-correlations instead of the signals themselves are back-propagated. Using the CINT algorithm, we can cancel some parts of the random phases present in the signals recorded at the receivers and arising from the multiple scattering of the acoustic waves on the inhomogeneities of the medium. The influence of the random medium can be mitigated as can be seen on \fref{fig:resnoflow} where CINT images are relatively stable and free of noise. Nevertheless the cross-range and range resolutions are slightly altered, as expected. The peaks centered at each reflector are rather blurry here, as a result of the smoothing of images induced by the CINT algorithm. Forming good images is based on a trade-off between the smoothing necessary to mitigate the noise, and the blurring that it implies. This trade-off is obtained by a relevant choice of the window parameters $\smash{\FWd}$ and $\smash{\SWd}$ of the CINT imaging function \eqref{eq:CINT-reflector}. The choice of an optimal window frequency $\smash{\FWd}$ contributes to enhance range resolution, while the choice of an optimal window length $\smash{\SWd}$ contributes to enhance cross-range resolution. \alert{These parameters can be determined by a statistical analysis of the data (using variograms for example), by the adaptive algorithm of \citep{BOR06b} which uses an optimization criterion based on the quality of the image as it is formed, or by a proper a priori choice. Here they have been selected incrementally by tests and trials in order to obtain images of good apparent resolutions with the least blurring. In \fref{fig:resnoflow} the ``optimal" choice $\smash{\FWd} = 0.1 B$ and $\smash{\SWd}= 0.8  a$ has been used. The theoretical range and cross-range resolutions of CINT in pixel units are then $2\smash{f_\iref}/\smash{\FWd}\simeq 13$ and $2L/\smash{\SWd}\simeq2$, respectively. We thus consider that the actual resolution achieved in \fref{fig:resnoflow} validates the CINT algorithm in quiescent media for its extension to moving media in the next \sref{sec:moving-random-medium}}.

\section{Imaging in a moving random medium}\label{sec:moving-random-medium}

In this section, we consider again the \alert{active imaging} configuration depicted in \fref{fg:ImgConfig} where the ambient medium is now moving at the velocity $\smash{\vref}$ which may be either uniform, or randomly perturbed around its 	average uniform value. We first assume that the speed of sound is constant in order to highlight the role of the Doppler compensation factor \eqref{eq:compensation} for the KM and CINT imaging algorithms in the moving medium. The speed of sound is also randomly perturbed around its average value, as in the foregoing section. \alert{The range of the three reflectors is again $L=90\wavelref$, where $\wavelref$ is the typical wavelength of the probing signals. }

\subsection{Active imaging in a random medium moving at constant velocity}\label{subsec:Rand:Med:Hom:Spe}

\begin{figure}[ht!]
\centering
\subfigure[Constant speed of sound]{\includegraphics[scale = 0.45]{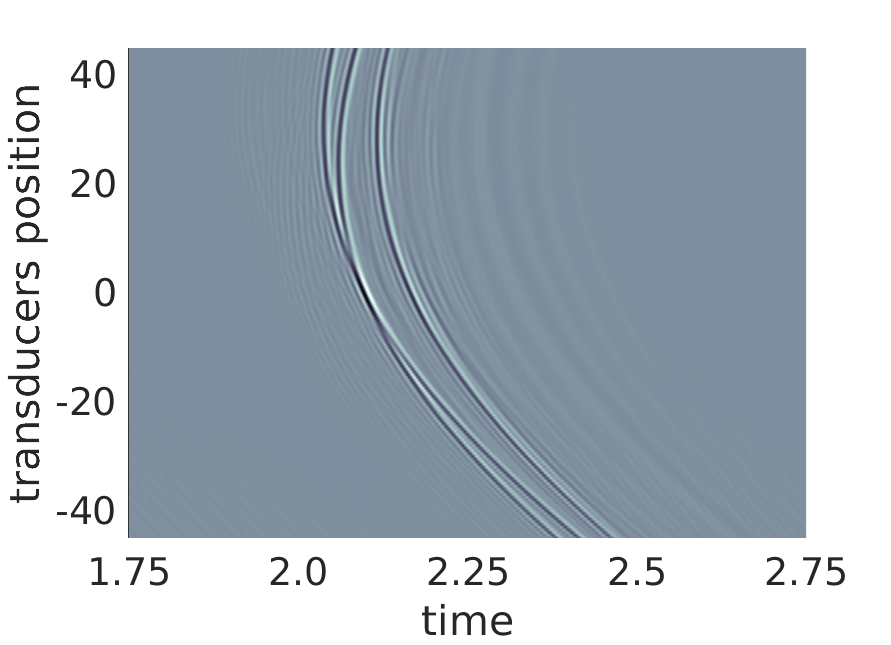}}
\subfigure[Random speed of sound]{\includegraphics[scale = 0.45]{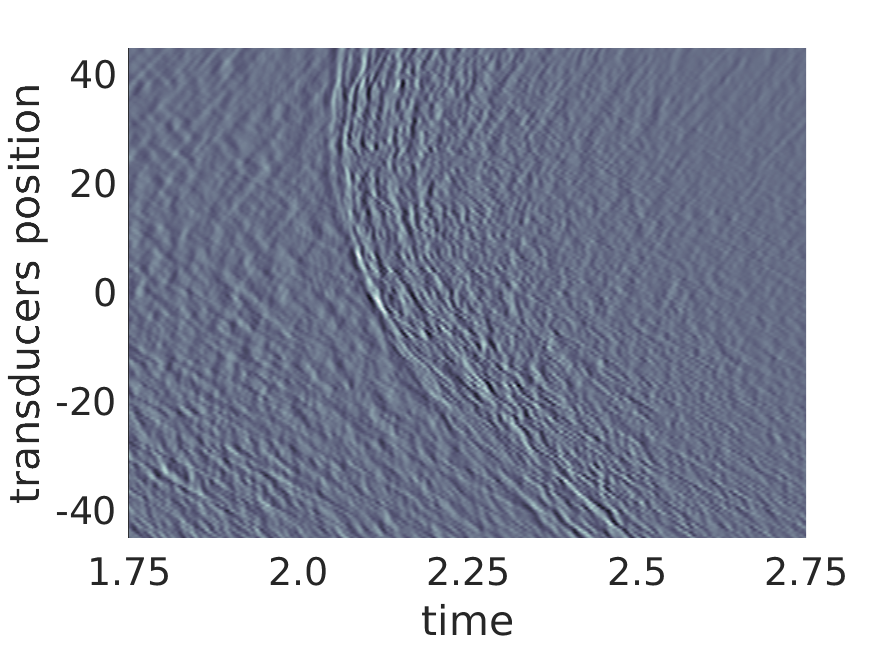}}
\caption{Active imaging in a moving medium in the configuration of \fref{fg:ImgConfig} with constant velocity $\Mach=0.3$ in the cross-range upward direction. (a) Time traces recorded at the receivers for a constant speed of sound $\smash{\celm}$. (b) Time traces recorded at the receivers for a random speed of sound with average $\smash{\celm}$ and standard deviation $\sigma=3\%$. The horizontal axis is time (\alert{in units of $L/\celm$}) and the vertical axis is the array transducer location (\alert{in units of $\smash{\wavelref}$}).}\label{fig:sig-fh}
\end{figure}

The ambient medium is moving at a constant velocity $\vref=\celm\times(0,\Mach)$ in the cross-range upward direction with Mach number $\Mach=0.3$ and average speed of sound $\smash{\celm}$. We use the KM imaging function \eqref{eq:Kirchhoff-in-flow:actif} and CINT imaging function \eqref{eq:CINT-in-folw-reflector} with $N_{\rm r}=N=185$ and $N_{\rm s}=1$ to localize the three fixed reflectors in \fref{fg:ImgConfig} (active  imaging configuration). The central transducer of the array is again used as the single source with the emitted signal equal to (\ref{eq:pulse}). Numerically generated data recorded at the array are shown in \fref{fig:sig-fh} for either a constant or a random speed of sound. Because the ambient medium is now moving upward in the cross-range direction, we note that the symmetry with respect to the transducer used as a source is lost compared to \fref{fg:sig-nofh}. We thus understand that this phenomenon has to be compensated by a correction factor in our imaging functions. This is the role of the Doppler compensation factor $\smash{\Doppler}$ of \eref{eq:compensation}. In \fref{fig:sig-fh}, the echoes from the reflectors are still observable. Thus if the compensation factor is correct, we should be able to localize each reflector at least for a constant speed of sound.

\begin{figure}[ht!]
\centering
\subfigure[KM image without $\smash{\Doppler}$]{\includegraphics[scale = 0.45]{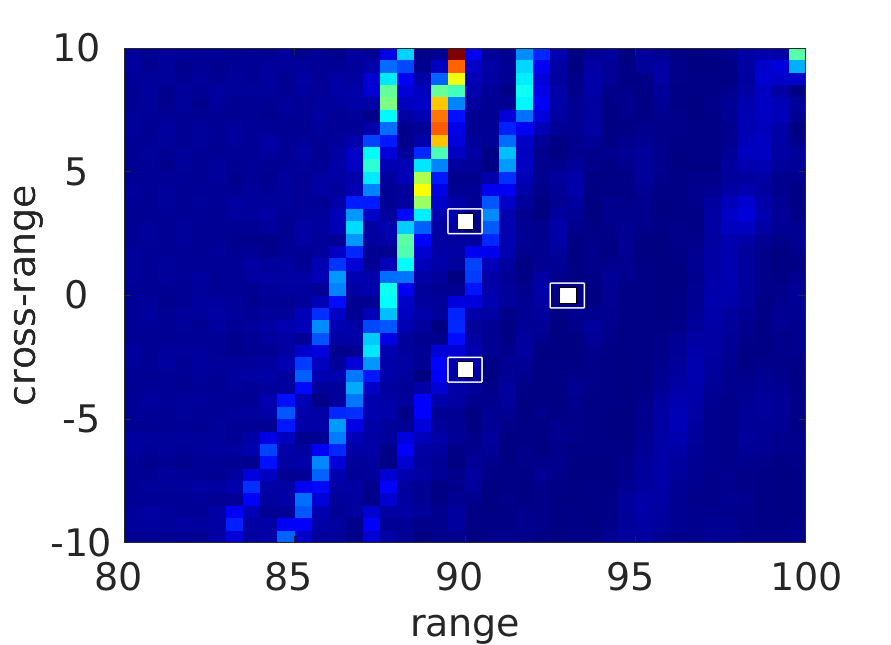}}
\subfigure[KM image with $\smash{\Doppler}$]{\includegraphics[scale = 0.45]{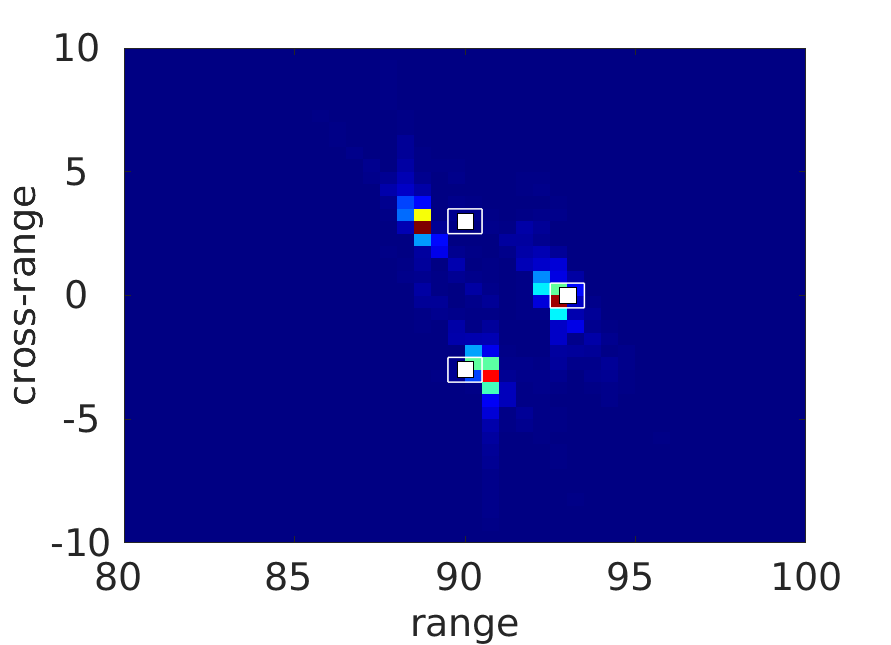}}
\caption{KM active imaging in a moving medium in the configuration of \fref{fg:ImgConfig} with constant velocity $\Mach=0.3$ in the cross-range upward direction and constant speed of sound $\smash{\celm}$. Influence of the Doppler compensation factor $\smash{\Doppler}$: (a) KM image \eqref{eq:Kirchhoff-source:actif} without $\smash{\Doppler}$; (b) KM image \eqref{eq:Kirchhoff-in-flow:actif} with $\Doppler$. The reflectors to be localized are shown by white squares {\tiny$\square$}. Range and cross-range dimensions are in units of $\wavelref$.}\label{fg:KIRC-gamma-unif}
\end{figure}

In order to highlight the role of the Doppler compensation factor, we compare in \fref{fg:KIRC-gamma-unif} the KM imaging function \eqref{eq:Kirchhoff-source:actif} (without $\smash{\Doppler}$, left) with the KM imaging function \eqref{eq:Kirchhoff-in-flow:actif} (with $\smash{\Doppler}$, right). The moving medium has a constant speed of sound and the search domain is a square of size $\smash{20\wavelref\times20\wavelref}$ centered around $(90\wavelref,0)$. When $\smash{\Doppler}$ is ignored, the positions of the three reflectors cannot be identified at all since the signals are back-propagated in a fictitious medium at rest. Conversely, when $\smash{\Doppler}$ is taken into account in the KM imaging function, these positions can be identified relatively well. Nevertheless, we notice that there is a slight shift for the top reflector. There can be several explanations: discretization errors (mesh, solver), imperfect reflections of the waves on the reflectors
 and/or imperfect absorption on the edges of the computational domain, 
or most likely errors induced in the model by the compensation factor $\smash{\Doppler}$ which is only a first-order corrector. From now on, $\smash{\Doppler}$ will be considered to form all KM and CINT images.

\begin{figure}[ht!]
\centering
\subfigure[KM image in random medium \#1]{\includegraphics[scale = 0.45]{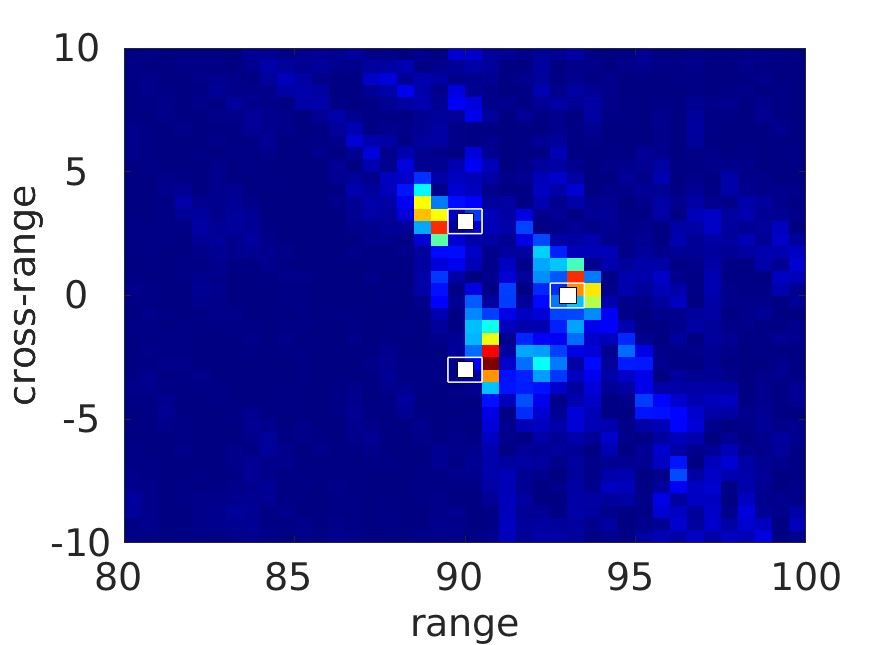}}
\subfigure[KM image in random medium \#2]{\includegraphics[scale = 0.45]{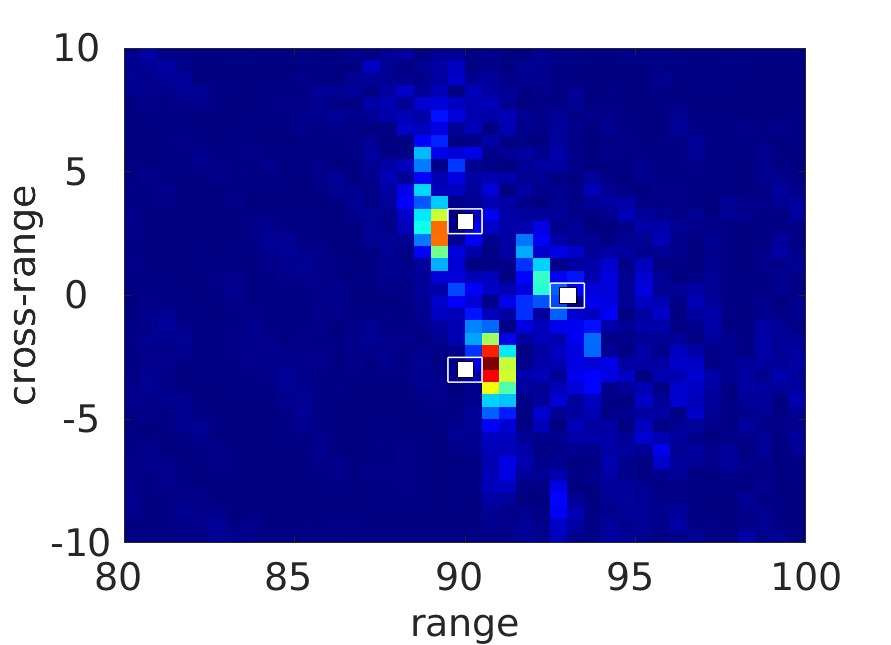}}\\
\subfigure[CINT image in random medium \#1]{\includegraphics[scale = 0.45]{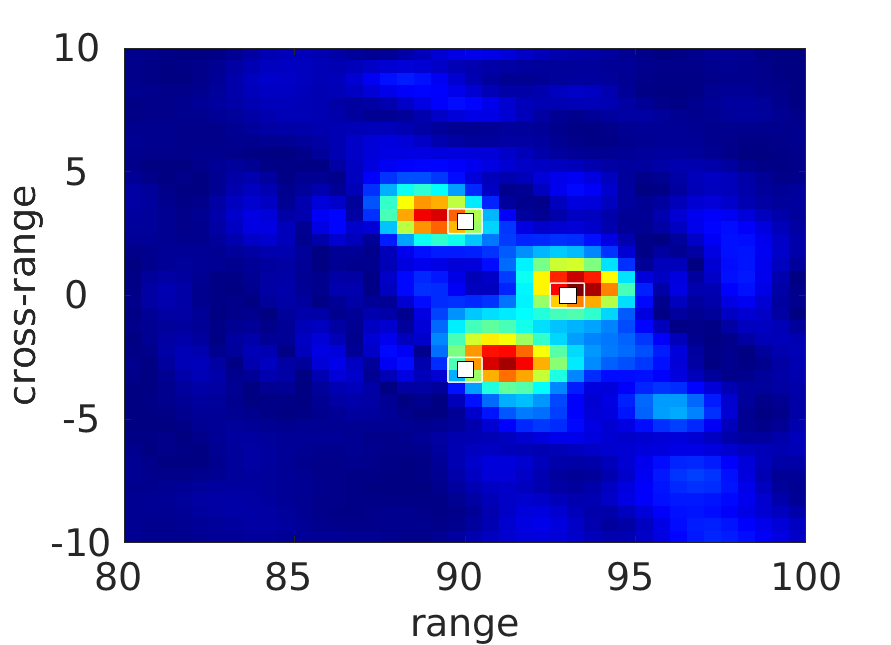}}
\subfigure[CINT image in random medium \#2]{\includegraphics[scale = 0.45]{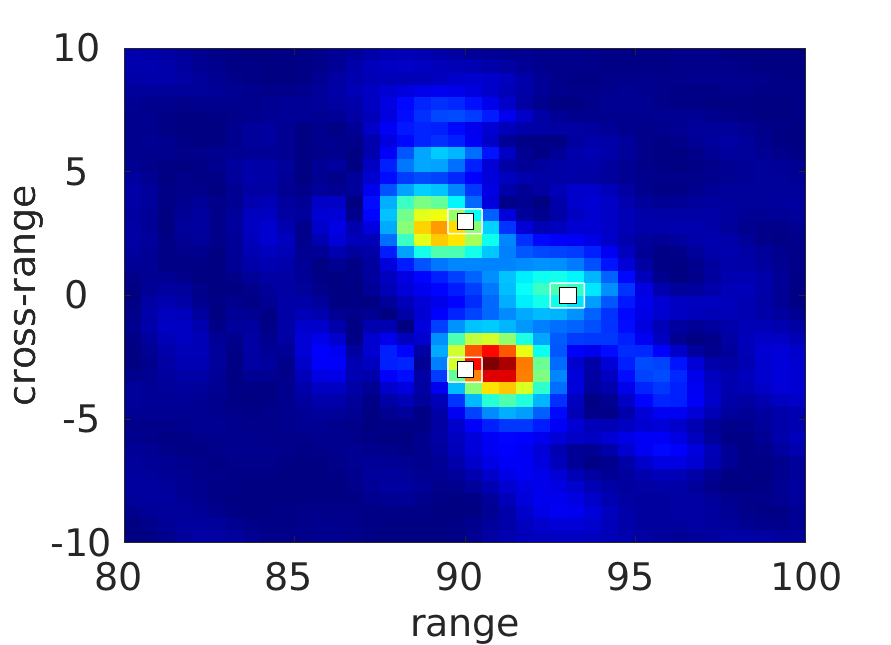}}
\caption{Active imaging in a moving medium in the configuration of \fref{fg:ImgConfig} with constant velocity $\Mach=0.3$ in the cross-range upward direction and random speed of sound with average $\smash{\celm}$ and standard deviation $\sigma=3\%$. Comparison of KM and CINT images for two realizations of the random medium: (a) Squared KM image for the first realization of the medium; (b) Squared KM image for the second realization of the medium; (c) CINT image for the first realization of the medium; (d) CINT image for the second realization of the medium. The CINT parameters $\smash{\FWd} = \smash{B/12}$ and $\smash{\SWd} = 0.9 a$ are used for both media. The reflectors to be localized are shown by white squares {\tiny$\square$}. Range and cross-range dimensions are in units of $\wavelref$.}\label{fg:CINT-finh}
\end{figure}

Now the ambient medium moving at the constant velocity $\Mach=0.3$ in the cross-range upward direction has a random speed of sound with average $\celm$ and standard deviation $\sigma=3\%$. The KM imaging function for two realizations of the medium is shown in \fref{fg:CINT-finh}. As in \fref{fig:resnoflow}, we observe noisy images that differ from one realization to another. We thus form the CINT imaging function \eqref{eq:CINT-in-folw-reflector} from the signals recorded by the receivers. The values of the parameters $\smash{\FWd}$ and $\smash{\SWd}$ are found by successive tests as in the previous experiments of \sref{sec:Num:res}. The results can be seen in \fref{fg:CINT-finh} alike. \alert{The theoretical range and cross-range resolutions of CINT in pixel units are $2\smash{f_\iref}/\smash{\FWd}\simeq 16$ and $2L/\smash{\SWd}\simeq2$, respectively}.
We show the results obtained with two realizations of the random medium and we observe the same behaviors as in \sref{sec:Num:res}: the KM algorithm is statistically unstable while the CINT algorithm is statistically stable. 
For the KM function of \fref{fg:CINT-finh} (top), we have difficulties finding the positions of the three reflectors since the image is noisy. For the CINT function of \fref{fg:CINT-finh} (bottom), the positions of the three reflectors appear much more clearly. The uncertainties present on the image obtained by the KM algorithm have practically vanished. Nevertheless, we have some blurring that comes from the smoothing induced by the CINT algorithm, as expected. These images could be improved by the adaptive algorithm of \citep{BOR06b} though. 

\subsection{Active imaging in a random medium moving at random velocity}\label{subsec:Rand:Med:Rand:Spe}

The ambient medium is now moving at a random velocity $\smash{\vref}=\celm\times(\sigma\mu_1(\rv),\Mach(1+\sigma\mu_2(\rv)))$ in the cross-range upward direction with Mach number $\Mach = 0.3$, \alert{average speed of sound $\smash{\celm}$}, and standard deviation $\sigma=3\%$. The disturbances $(\smash{\mu_1}(\rv),\,\rv\in\Rset^2)$ and $(\smash{\mu_2}(\rv),\,\rv\in\Rset^2)$ are stationary, mean-zero and independent second-order random processes. Gaussian covariance functions \eqref{eq:Markov:cor} are chosen for both of them with identical correlation lengths. These perturbations are simulated by random Fourier series of stationary processes \citep{POI89,SHI71}. 
We note that this random ambient velocity $\smash{\vref}$ does not necessarily solve the ambient flow equations \eqref{eq:stat_flow} and the covariance functions of its turbulent components do not fulfill the necessary condition stated in \citep{BAT53,VON38,ROB40}. This simplified model is a first attempt to take into account turbulent phenomena with the KM and CINT imaging algorithms. More realistic models shall be envisaged in future works.

\begin{figure}[ht!]
\centering
\subfigure[Constant speed of sound]{\includegraphics[scale = 0.45]{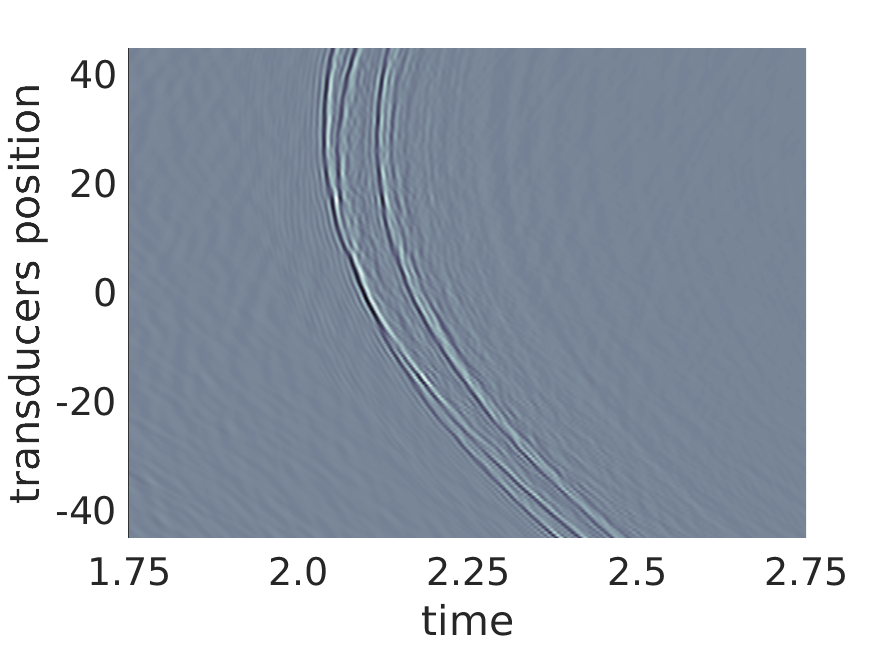}}
\subfigure[Random speed of sound]{\includegraphics[scale = 0.45]{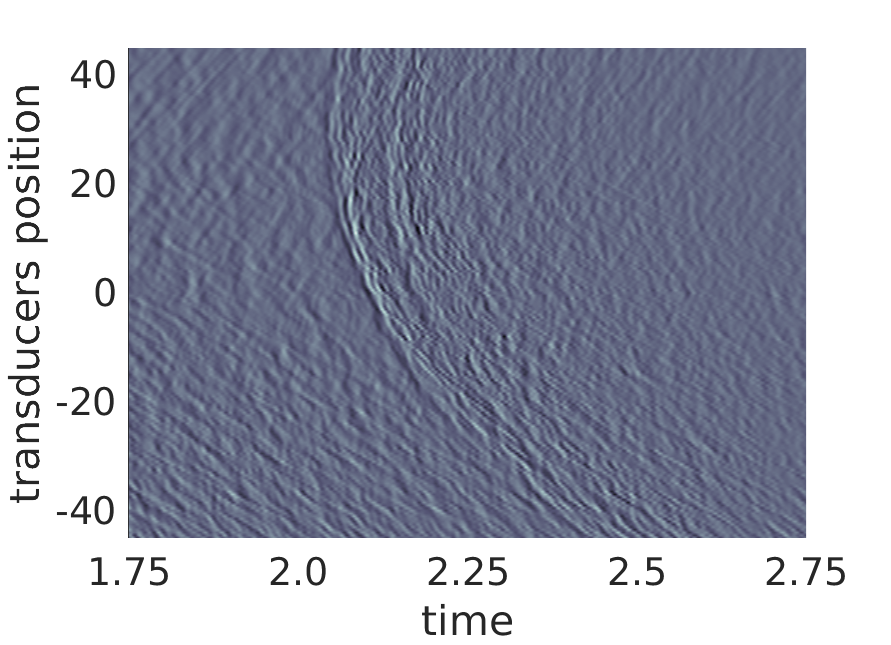}}
\caption{Active imaging in a moving medium in the configuration of \fref{fg:ImgConfig} with random velocity with average $\Mach=0.3$ in the cross-range upward direction. (a) Time traces recorded at the receivers for a constant speed of sound $\smash{\celm}$. (b) Time traces recorded at the receivers for a random speed of sound with average $\smash{\celm}$ and standard deviation $\sigma=3\%$. The horizontal axis is time (\alert{in units of $L/\smash{\celm}$}) and the vertical axis is the array transducer location (\alert{in units of $\smash{\wavelref}$}).}\label{fig:sig-fnoisehom}
\end{figure}

\begin{figure}[ht!]
\centering
\subfigure[KM image without $\smash{\Doppler}$]{\includegraphics[scale = 0.45]{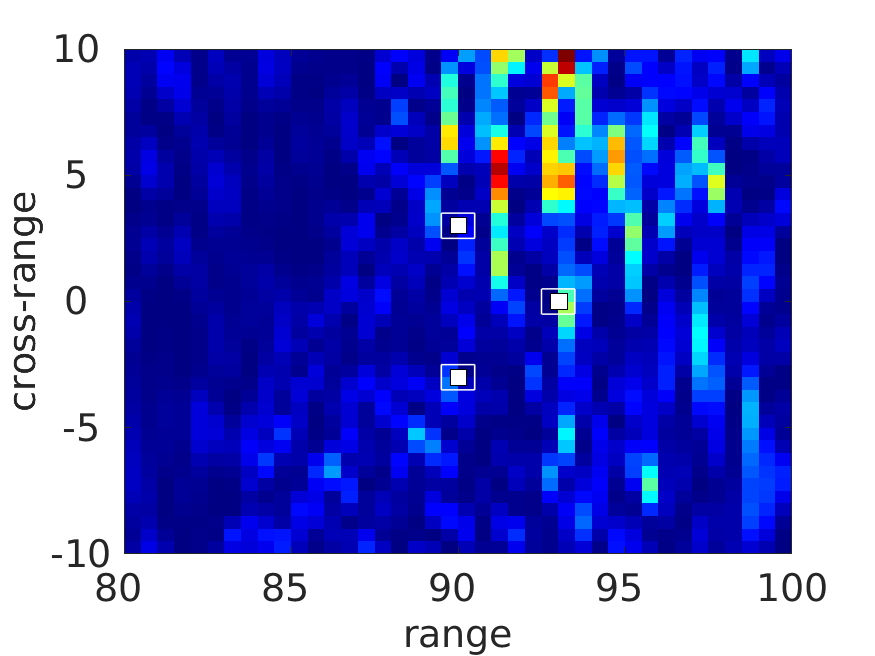}}
\subfigure[KM image with $\smash{\Doppler}$]{\includegraphics[scale = 0.45]{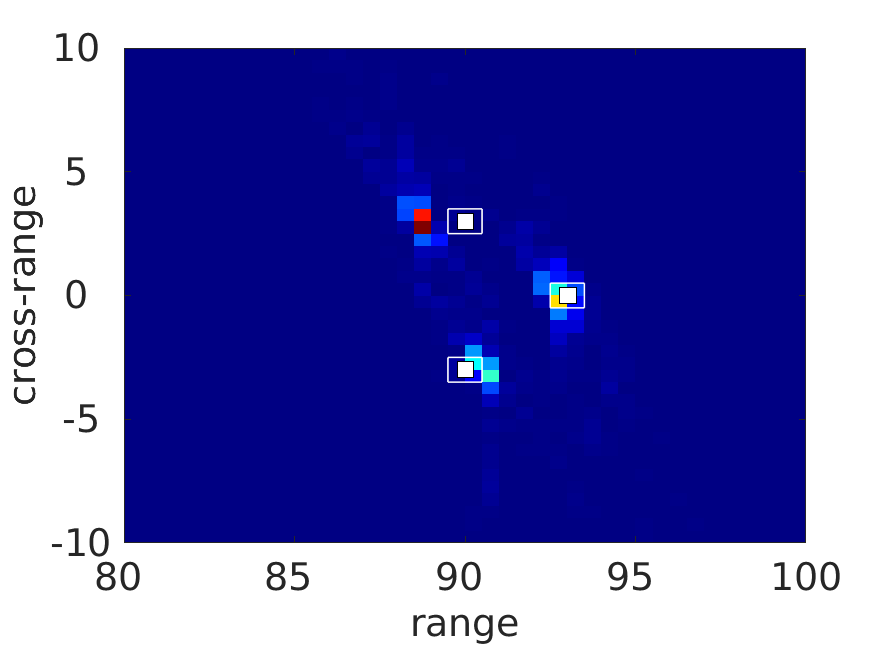}}
\caption{KM active imaging in a moving medium in the configuration of \fref{fg:ImgConfig} with random velocity with average $\Mach=0.3$ in the cross-range upward direction and constant speed of sound $\smash{\celm}$. Influence of the Doppler compensation factor $\smash{\Doppler}$: (a) KM image \eqref{eq:Kirchhoff-source:actif} without $\smash{\Doppler}$; (b) KM image \eqref{eq:Kirchhoff-in-flow:actif} with $\Doppler$. The reflectors to be localized are shown by white squares {\tiny$\square$}. Range and cross-range dimensions are in units of $\wavelref$.}\label{fig:KIRC-hfnoise}
\end{figure}

Numerically generated data recorded at the array are shown in \fref{fig:sig-fnoisehom} for either a constant or a random speed of sound. Again, the symmetry with respect to the transducer used as a source--the central one--is lost. Compared to the moving medium with constant velocity, we observe here echoes fairly close to the echoes in \fref{fig:sig-fh} but noisier. This is induced by the velocity of the ambient medium which is now random. Thus, the waves propagating in this medium will also be noisier. Even in the presence of this noise, we still see the different fronts representing the echoes from the three reflectors. The role of the Doppler compensation factor $\smash{\Doppler}$ is also highlighted in \fref{fig:KIRC-hfnoise}. Here the KM imaging function \eqref{eq:Kirchhoff-source:actif} (without $\smash{\Doppler}$, left) and the KM imaging function \eqref{eq:Kirchhoff-in-flow:actif} (with $\smash{\Doppler}$, right) for one realization of the random velocity $\smash{\vref}$ of the ambient medium are shown. The compensation factor is computed for the average velocity $\celm\times(0,\Mach)$ of the ambient medium. The search domain is again a square of size $\smash{20\wavelref\times20\wavelref}$ centered around $(90\wavelref,0)$. The KM algorithm accounting for the Doppler compensation factor works well for this configuration (constant speed of sound) even if the ambient velocity is random.  

\begin{figure}[ht!]
\centering
\subfigure[KM image in random medium \#1]{\includegraphics[scale = 0.45]{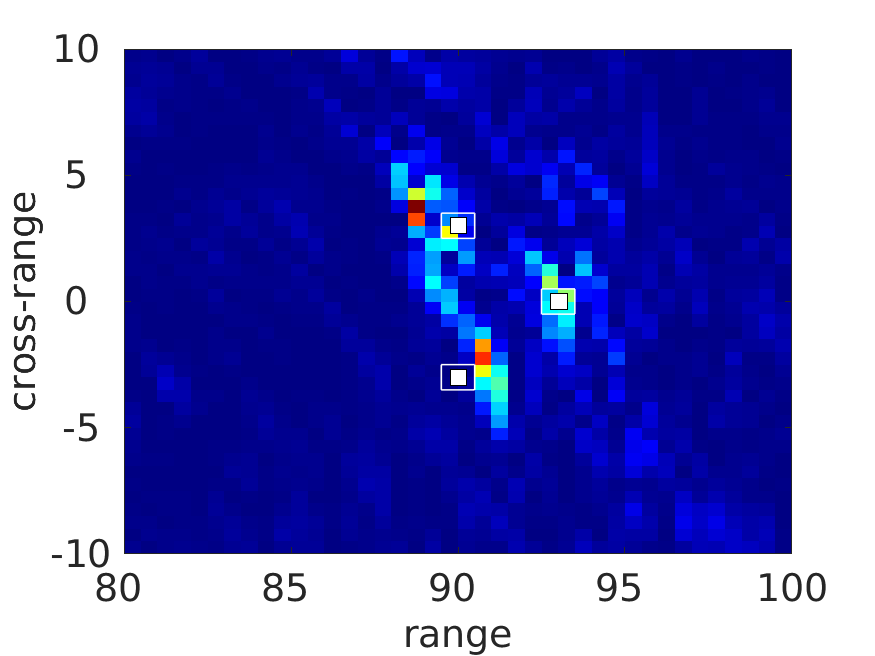}}
\subfigure[KM image in random medium \#2]{\includegraphics[scale = 0.45]{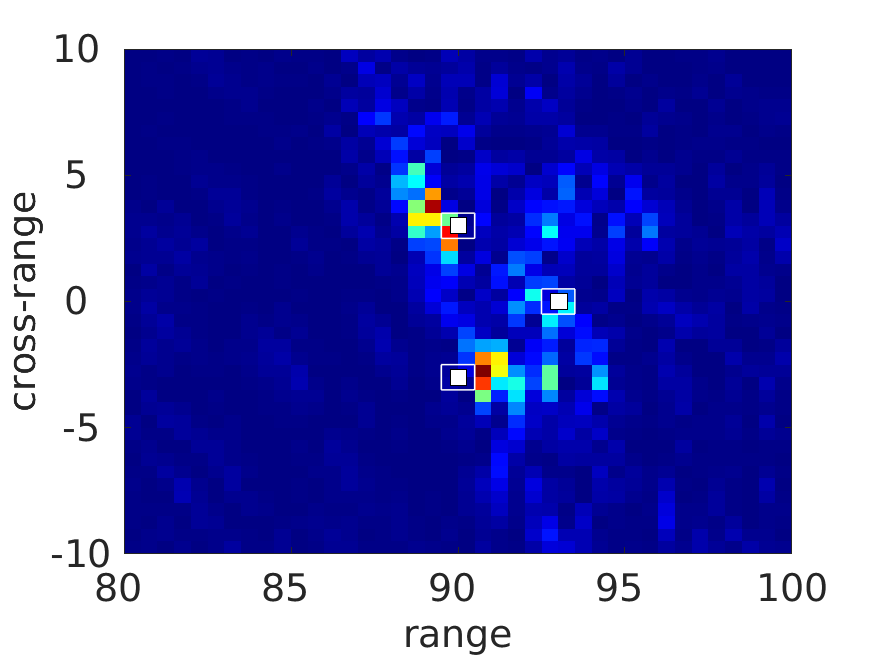}}
\subfigure[CINT image in random medium \#1]{\includegraphics[scale = 0.45]{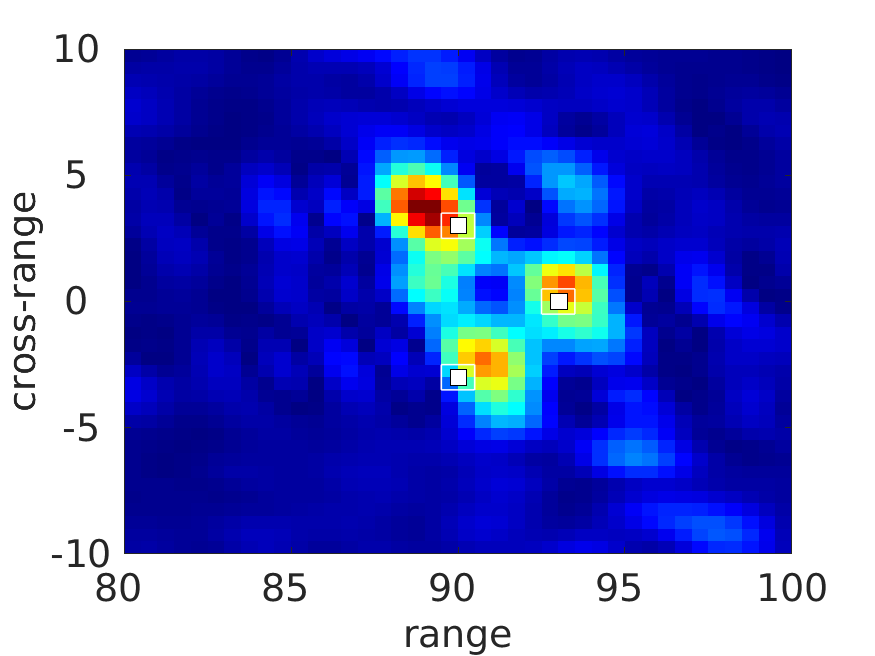}}
\subfigure[CINT image in random medium \#2]{\includegraphics[scale = 0.45]{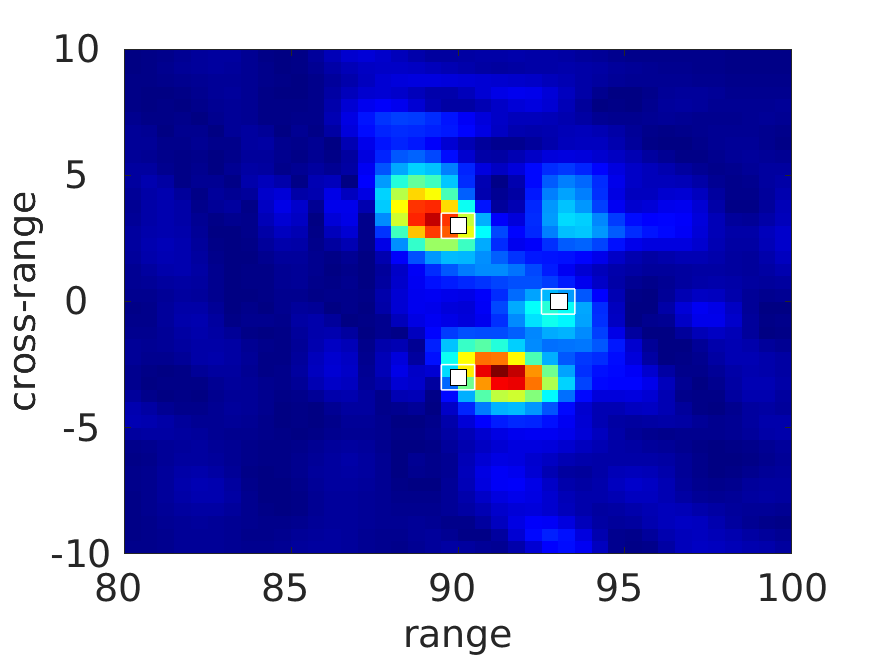}}
\caption{Active imaging in a moving medium in the configuration of \fref{fg:ImgConfig} with random velocity with average $\Mach=0.3$ in the cross-range upward direction and random speed of sound with average $\smash{\celm}$ and standard deviation $\sigma=3\%$. Comparison of KM and CINT images for two realizations of the random medium: (a) Squared KM image for the first realization of the medium; (b) Squared KM image for the second realization of the medium; (c) CINT image for the first realization of the medium; (d) CINT image for the second realization of the medium. The CINT parameters $\smash{\FWd} = 0.09 B$ and $\smash{\SWd} = 0.9 a$ are used for both media. The reflectors to be localized are shown by white squares {\tiny$\square$}. Range and cross-range dimensions are in units of $\wavelref$.}\label{fig:res-inhflownoise}
\end{figure}

Images obtained by the KM algorithm for two realizations of the ambient medium with random velocity and random speed of sound are shown in \fref{fig:res-inhflownoise}. Again, we observe the statistical instability of the KM functions, since the positions of the reflectors can not be determined unambiguously for any realization of the medium. We thus form the CINT imaging function \eqref{eq:CINT-in-folw-reflector} from the signals recorded by the receivers. The parameters $\smash{\FWd}$ and $\smash{\SWd}$ are chosen by successive tests as in the previous experiments. The results can be seen in \fref{fig:res-inhflownoise} alike. \alert{The theoretical range and cross-range resolutions of CINT in pixel units are $2\smash{f_\iref}/\smash{\FWd}\simeq 15$ and $2L/\smash{\SWd}\simeq2$, respectively}.
We  show the results obtained with two realizations of the medium and we can make the same observations as in \sref{sec:Num:res}. 
Using the KM function (\fref{fig:res-inhflownoise}, top), the positions of the three reflectors can not be distinguished. On the other hand, using the CINT function (\fref{fig:res-inhflownoise}, bottom), the positions of the three reflectors can be distinguished more easily. Two issues are worth noticing though. First, a loss in resolution due to smoothing is observed. Second, slight offsets of the positions of the reflectors (of the order of a few pixels) are observed. This phenomenon can be explained by the fact that the Doppler compensation factor $\smash{\Doppler}$ is a first-order correction term.

\section{Passive imaging through a synthetic turbulent jet flow} \label{sec:Jet}

\newcommand{\markerone}{\raisebox{0.pt}{\tikz{\filldraw[fill=gray, draw=black] (0,0) rectangle (0.3,0.2);}}}
\newcommand{\markertwo}{\raisebox{2pt}{\tikz{\draw[draw=red] (0,0) -- (0.5,0);}}}
\newcommand{\markerthree}{\raisebox{2pt}{\tikz{\draw[draw=blue] (0,0) -- (0.5,0);}}}

\begin{figure}[ht!]
\centering
\includegraphics[scale=1.5]{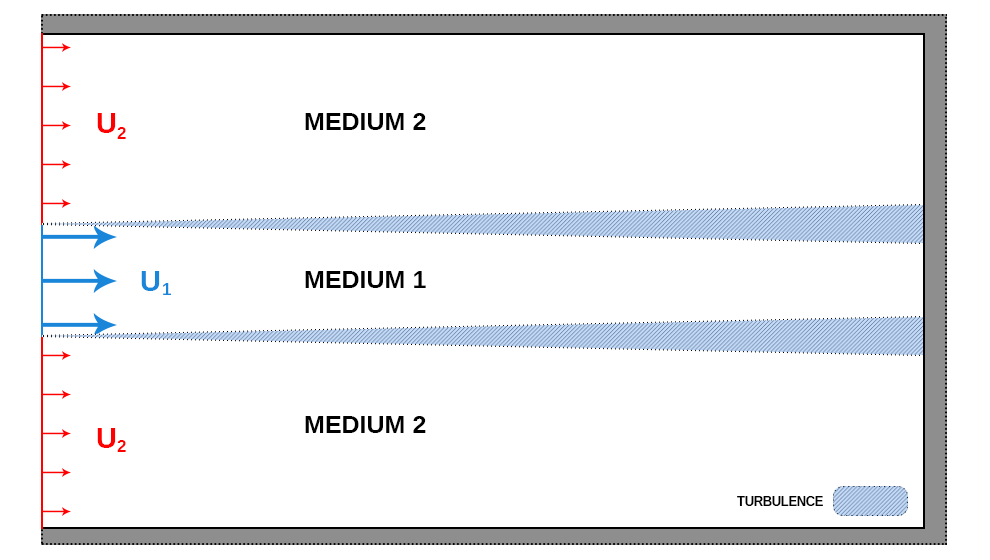}
\caption{Synthetic jet flow configuration: \protect\markerone\ non-reflection Pad\'e boundary conditions; \protect\markerthree\ jet inflow boundary condition for the central zone with $\smash{U_1} = 100$ m/s;  \protect\markertwo\ jet inflow boundary condition outside the central zone with $\smash{U_2} = 50$ m/s.}\label{fg:SetUpJet}
\end{figure}

In this section we finally aim at imaging sources through a synthetic turbulent jet flow using the KM and CINT algorithms. A sketch of the configuration considered here is shown in \fref{fg:SetUpJet}. This case study is inspired by the experiments conducted by Candel \textit{et al.} in the seventies \citep{CAN75,CAN76a,CAN76b} at the low speed open wind tunnel of the Von K\'arm\'an institute, and more recently by Kr\"ober \textit{et al.} \citep{KRO13} at the Aeroacoustic Wind Tunnel Braunschweig (AWB) facility, and by Sijtsma \textit{et al.} \citep{SIJ14} at the DNW (the German-Dutch Wind Tunnels) PLST wind tunnel. The jet inflow boundary conditions on the left side of \fref{fg:SetUpJet} yield two sub-domains with average ambient flow velocities $\smash{\norm{\vref}=U_1}=100$ m/s and  $\smash{\norm{\vref}=U_2}=50$ m/s, respectively, with turbulent shear layers in between. Our goal is to image point sources lying in the first medium with an array of sensors placed in the second medium--that is, a passive imaging configuration in the terminology used in the foregoing sections.

\subsection{Synthetic jet model}\label{sec:jet:model}

The data for this generic jet configuration are synthesized numerically in two steps as follows:
\begin{itemize}
\item First, we simulate the ambient shear flow by solving the Euler equations \eqref{eq:Euler} using the DG solver SPACE \citep{DEL05,PEY17}; this is the CFD step. In order to generate turbulence, viscosity phenomenon must be present. By definition, Euler equations are applicable to adiabatic and inviscid flows. Nevertheless, some viscosity is actually generated by the numerical scheme. At first, we will use this numerical viscosity to simulate the turbulent shear layers.
\item Second, we simulate the propagation of acoustic waves in the ambient flow obtained in the CFD step by solving the linearized Euler equations \eqref{eq:LEE} using again the DG solver SPACE \citep{DEL05,PEY17}; this is the CAA step. In a first approximation, we will consider that the acoustic phenomena are much faster than the fluidic phenomena. As a result, the ambient flow will be considered as "frozen" (time independent). The flow obtained at the last time increment of the CFD step will be used as the carrier (ambient) flow for the CAA step.
\end{itemize}
Of course the ambient flow obtained by the foregoing computation is poorly resolved and cannot be considered as a realistic jet flow. We are actually not interested in computing precisely the turbulent structures for this jet configuration. Our aim is rather to test the KM and CINT imaging algorithms in such a heterogeneous medium. The turbulent layers as resolved by the present numerical scheme are sufficient to exhibit and discuss the main features of these algorithms. These complex structures will play the role of the "clutter" of the idealized experiments of \sref{sec:Num:res} and \sref{sec:moving-random-medium}. 

\subsection{Imaging setup}\label{sec:jet:config}

\begin{figure}[ht!]
\centering
\includegraphics[scale=1.5]{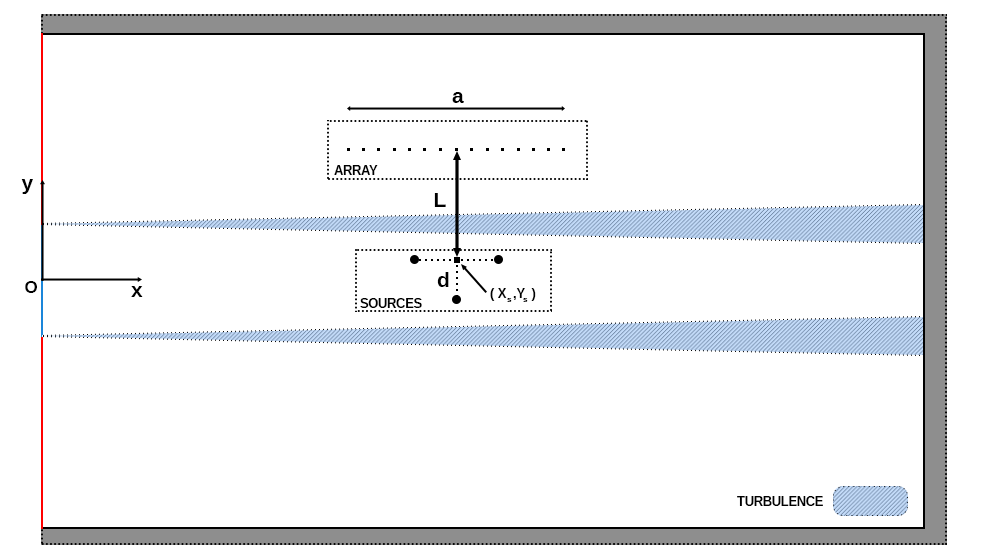}
\caption{Synthetic jet flow configuration: imaging setup. $a$ is the aperture of the sensors array, $L$ is the distance between the central sensor and the mid-point $\smash{(X_s,Y_s)}$ of the two upper sources which are $d$ apart from each other, and the lower source is also at the distance $d$ from the upper sources.}\label{fig:paramjet}
\end{figure}

A parameterization of the imaging setup for the synthetic jet flow of \fref{fg:SetUpJet} is shown in \fref{fig:paramjet}. We seek to determine the positions of three point sources of which emitted signals go through a turbulent shear layer. All sources emit the same signal given by \eref{eq:pulse}. The central frequency is here $\smash{f_\iref} = 50$ kHz. The fluid studied is the air defined by an average speed of sound $\celm = 340$ m/s at the temperature $T = 20 \degres$C. This corresponds to a wavelength $\wavelref = 6.8 \,10^{-3}$ m. The remote array contains $N=45$ sensors at a distance $ {\wavelref}/{2} $ from each other (a dense array).

\subsection{Flow computation}\label{sec:jet:calcul}

\begin{figure}[ht!]
\centering
\includegraphics[scale=0.4]{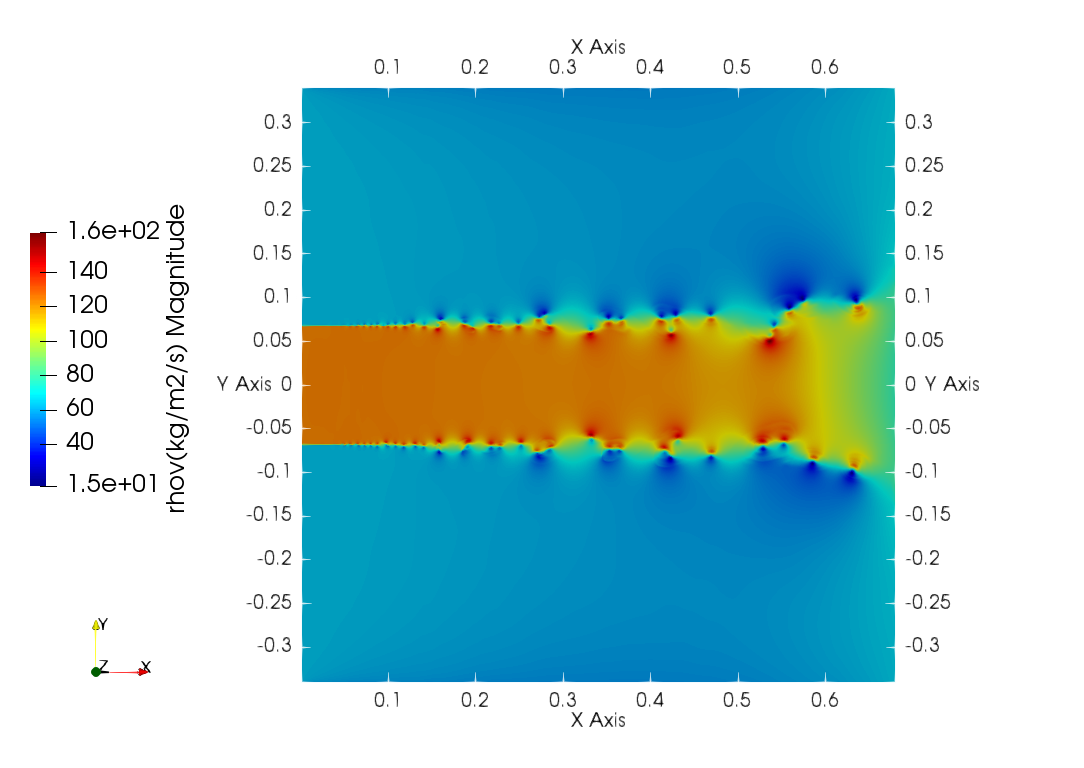}
\caption{Base flow obtained by solving the Euler equations (CFD step) using a DG finite element method for spatial discretization and a Runge-Kutta method for time discretization for the synthetic jet configuration of \fref{fg:SetUpJet}. The figure plots the norm of the momentum $\norm{\roi\vg}$.}\label{fg:flowJet}
\end{figure}

The same mesh is used for both the CFD and CAA steps. It is constructed so that there are 30 elements per wavelength $\wavelref$ (the central wavelength of the source) for a mesh size of $ 100 \wavelref \times 100 \wavelref $. We thus have $ 3000 \times 3000$ elements, that is a total of $9 \,10^6$ elements. The mesh is regular and constituted by four-node quadrilateral elements. The base flow computed in the CFD step is shown in \fref{fg:flowJet}. As a first approximation, we will consider it as frozen and use it as the ambient flow for testing the KM and CINT imaging algorithms. We first notice the development of a turbulent shear layer at the interfaces between the parts of the jet having different inflow velocities. This turbulent layer arises from the inflow velocity difference: $100$ m/s for the first medium located between $-10\wavelref$ and $10\wavelref$ along the range axis $\yj$ (the first medium), and $50$ m/s elsewhere (the second medium). 

\subsection{Refraction compensation mechanism}\label{sec:jet:refrac}

\begin{figure}[ht!]
\centering
\includegraphics[scale=0.3]{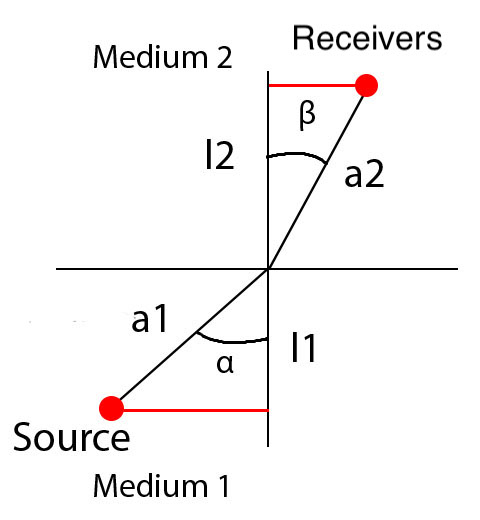}
\caption{Refracted acoustic wave at the interface between two media with speeds of sound $\smash{\cel_1}$ and $\smash{\cel_2}$: $\alpha$ is the angle of the incident wave, $\beta$ is  the angle of the refracted wave. The interface is the mean shear layer between the first and second media in \fref{fg:SetUpJet}.}\label{fig:setuprefraction}
\end{figure}

In addition to the Doppler compensation factor $\smash{\Doppler}$ of \eref{eq:compensation} which accounts for the convection effects, we must consider the fact that the acoustic waves generated by the source in the first medium will travel through media possibly with different speeds of sound; see \fref{fig:setuprefraction}. Therefore we introduce a refraction compensation mechanism in the imaging functions introduced in \sref{sec:func}. In this context the KM passive imaging function \eqref{eq:Kirchhoff-in-flow} and CINT passive imaging function \eqref{eq:CINT-in-flow} take the form:
\begin{equation}\label{eq:KM:refrac}
\begin{split}
\IFunction_\iKM(\rv^S) = \sum_{r=1}^N\pres\left(\rv_r,\gamma_{D_1}(\rv^S, \rv_\mathcal{O}, U_1\hat{\ev}_\xj)T_1 + \gamma_{D_2}(\rv_\mathcal{O},\rv_r, U_2\hat{\ev}_\xj) T_2\right)\,,
\end{split}
\end{equation}
and:
\begin{multline}\label{eq:CINT:refrac}
\IFunction_\iCINT(\rv^S;\FWd,\SWd)= \sum_{\scriptsize\begin{array}{c}q,r=1\\\norm{\rv_q-\rv_r}\leq\SWd\end{array}}^N\iint_{\norm{\om-\om'}\leq\FWd}\AutoCor_F(\rv_q,\rv_r,\om,\om') \\
\times \iexp^{-\ci\om(\gamma_{D_1}(\rv^S, \rv_\mathcal{O}, U_1\hat{\ev}_\xj) T_1 + \gamma_{D_2} (\rv_\mathcal{O},\rv_q, U_2\hat{\ev}_\xj)T_2)}\iexp^{\ci\om'(\gamma_{D_1}(\rv^S, \rv_{\mathcal{O}'}, U_1\hat{\ev}_\xj) T'_1 + \gamma_{D_2}(\rv_{\mathcal{O}'},\rv_r, U_2\hat{\ev}_\xj) T'_2)}\,\di\om\di\om'\,,
\end{multline}
respectively, where $\hat{\ev}_\xj$ is the unit vector along the (horizontal) direction of the ambient flow. The Doppler compensation factors $\smash{\gamma_{D_1}}$ and $\smash{\gamma_{D_2}}$ are calculated as in \eqref{eq:compensation} with speeds of sound $\smash{\cel_1}$ and $\smash{\cel_2}$ for the two media. Referring to \fref{fig:setuprefraction} where $\alpha$ is the angle of the incident wave, $\beta$ the angle of the refracted wave, and $\mathcal{O}$ is the point of the interface between both media where the wave is refracted, then the travel times within the first and second media are $\smash{T_1} = \smash{a_1/\cel_1}$ and $\smash{T_2} = \smash{a_2/\cel_2}$, respectively, where $\smash{a_1} = \smash{l_1/\cos\alpha}=\smash{\norm{\rv^S-\rv_\mathcal{O}}}$ and $\smash{a_2} = \smash{l_2/\cos\beta}=\smash{\norm{\rv_\mathcal{O}-\rv_q}}$, respectively (likewise, $\smash{T_1'} = \smash{a_1'/\cel_1}$ and $\smash{T_2} = \smash{a_2'/\cel_2}$ with $\smash{a_1'}=\smash{\norm{\rv^S-\rv_{\mathcal{O}'}}}$ and $\smash{a_2'}=\smash{\norm{\rv_{\mathcal{O}'}-\rv_r}}$). It is important to note that in this refraction compensation mechanism only the phase (travel times) is significant for localizing the source, and not the amplitude. This is the reason why no amplitude reflection/transmission coefficient is required in the imaging functions.

\subsection{Imaging of the sources}\label{sec:jet:Img}

\begin{figure}[ht!]
\centering
\subfigure[No compensation]{\includegraphics[scale = 0.22]{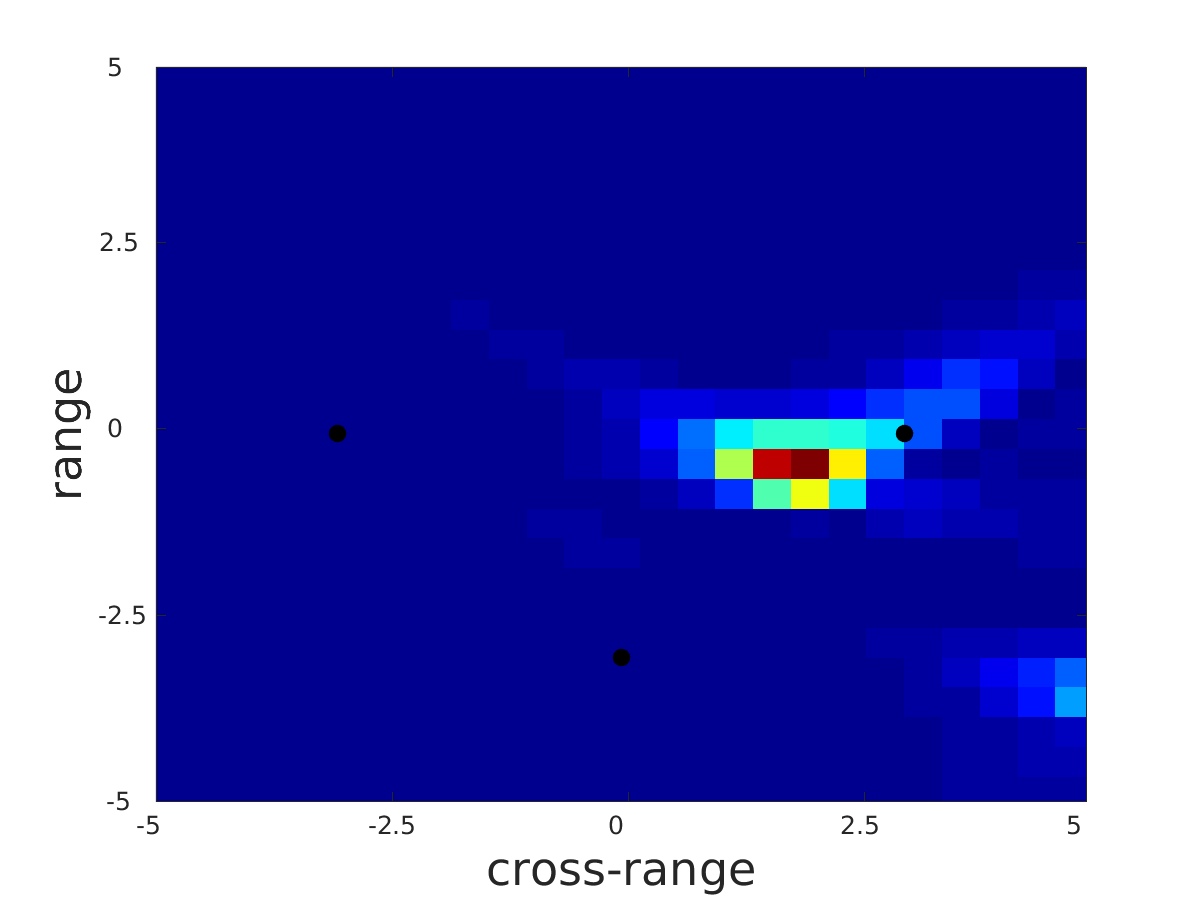}}
\subfigure[$\smash{\Doppler}$ compensation]{\includegraphics[scale = 0.22]{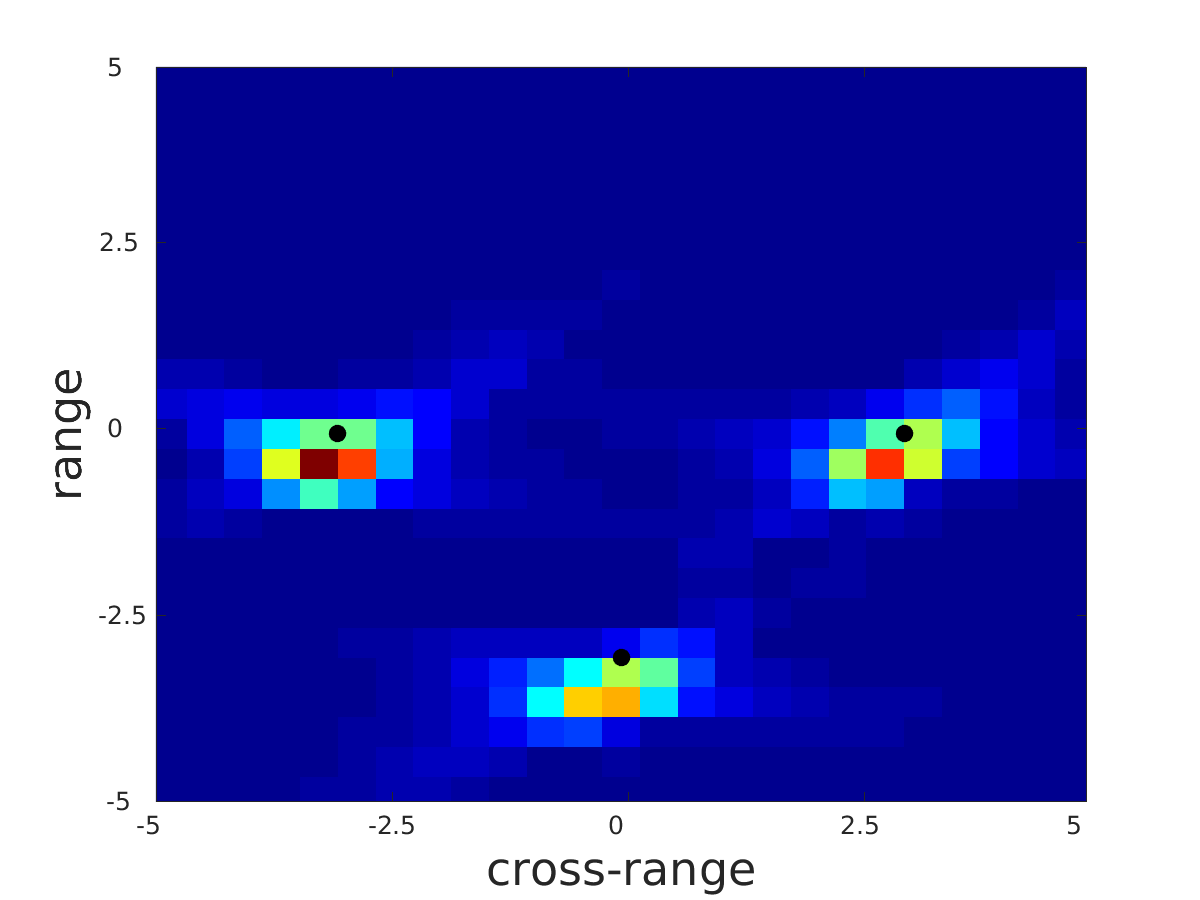}}
\subfigure[Full compensation]{\includegraphics[scale = 0.22]{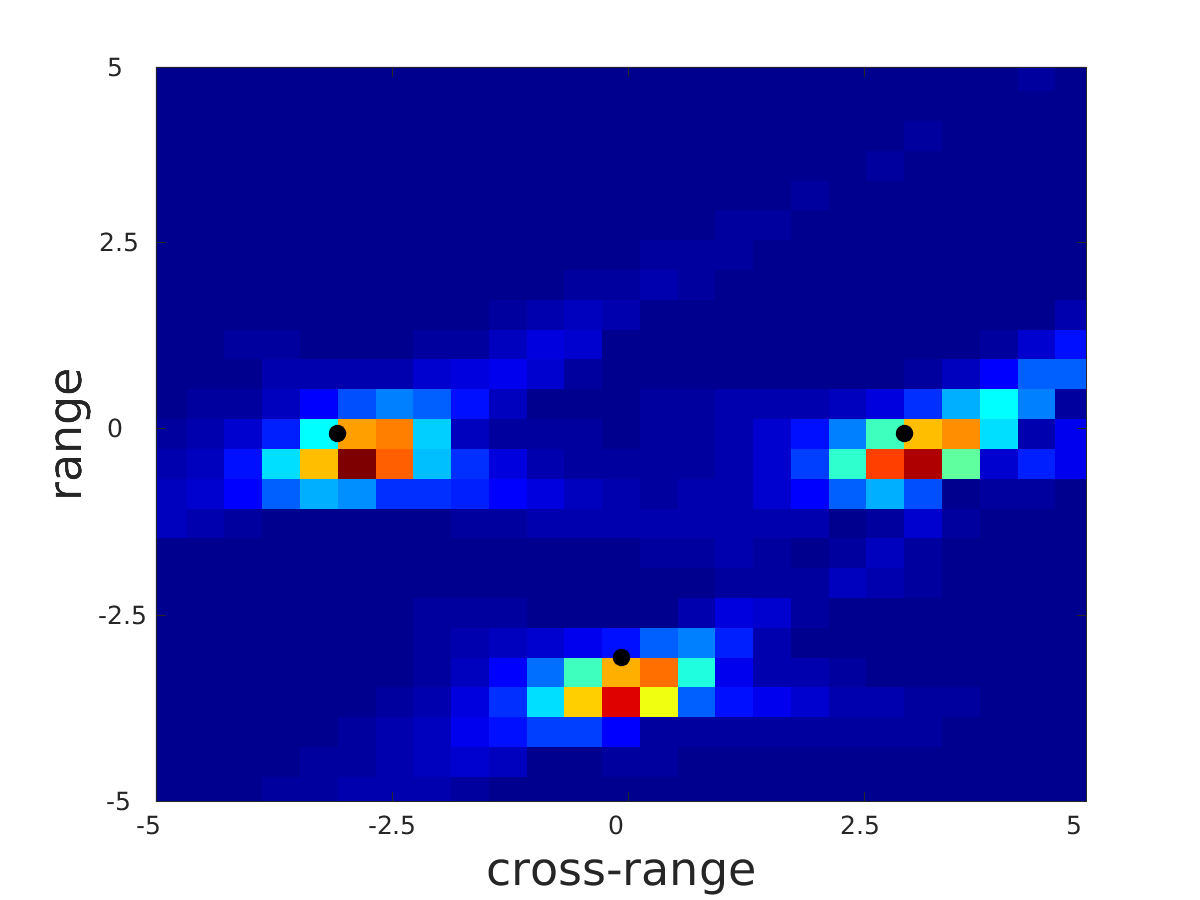}} \\
\subfigure[No compensation]{\includegraphics[scale = 0.22]{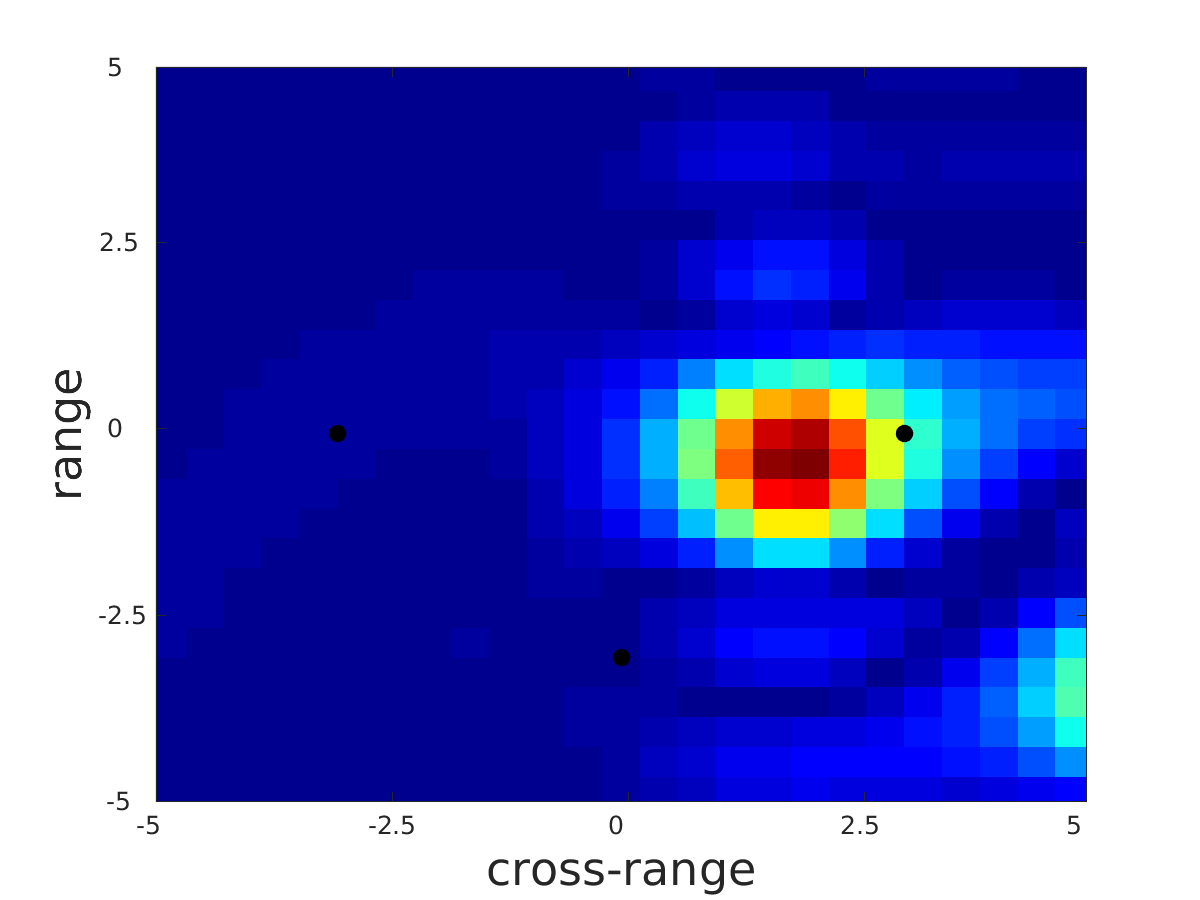}}
\subfigure[$\smash{\Doppler}$ compensation]{\includegraphics[scale = 0.22]{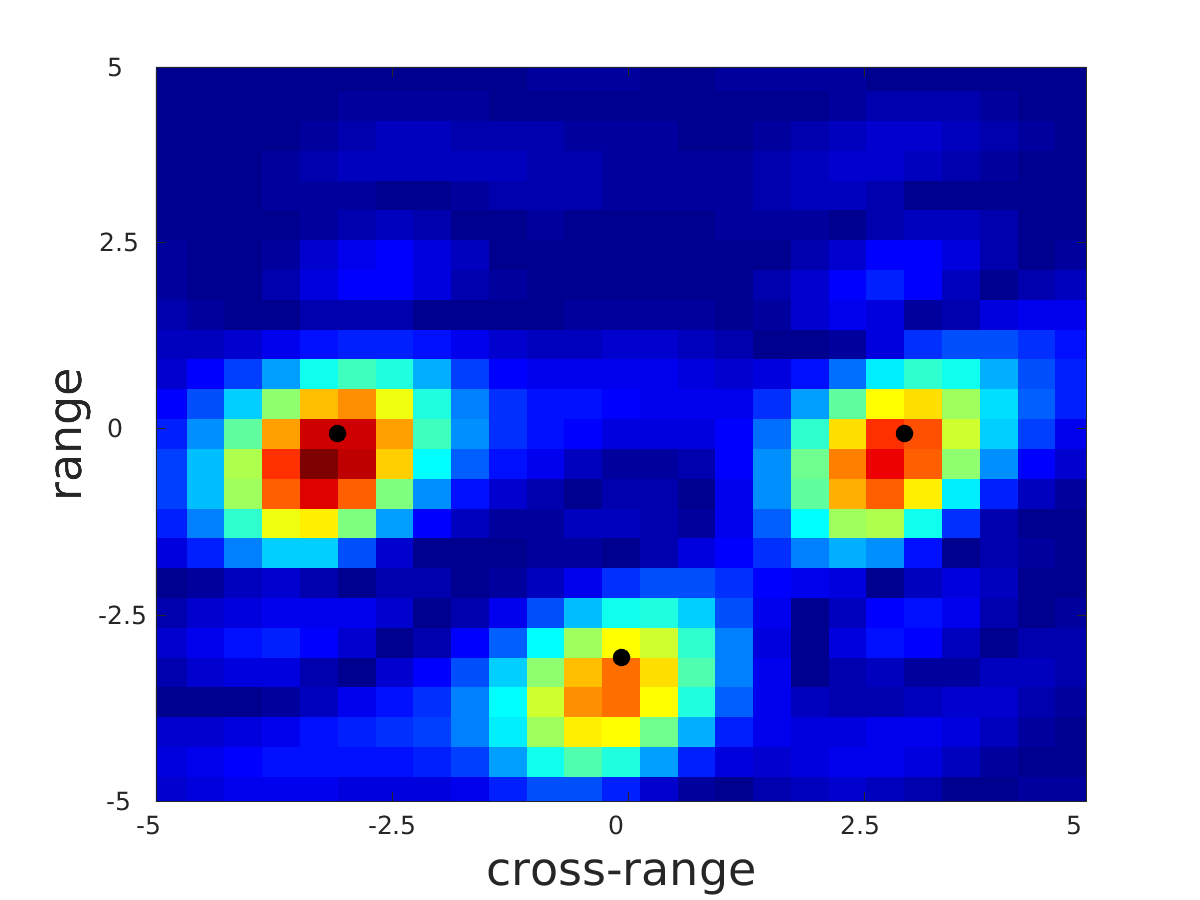}} 
\subfigure[Full compensation]{\includegraphics[scale = 0.22]{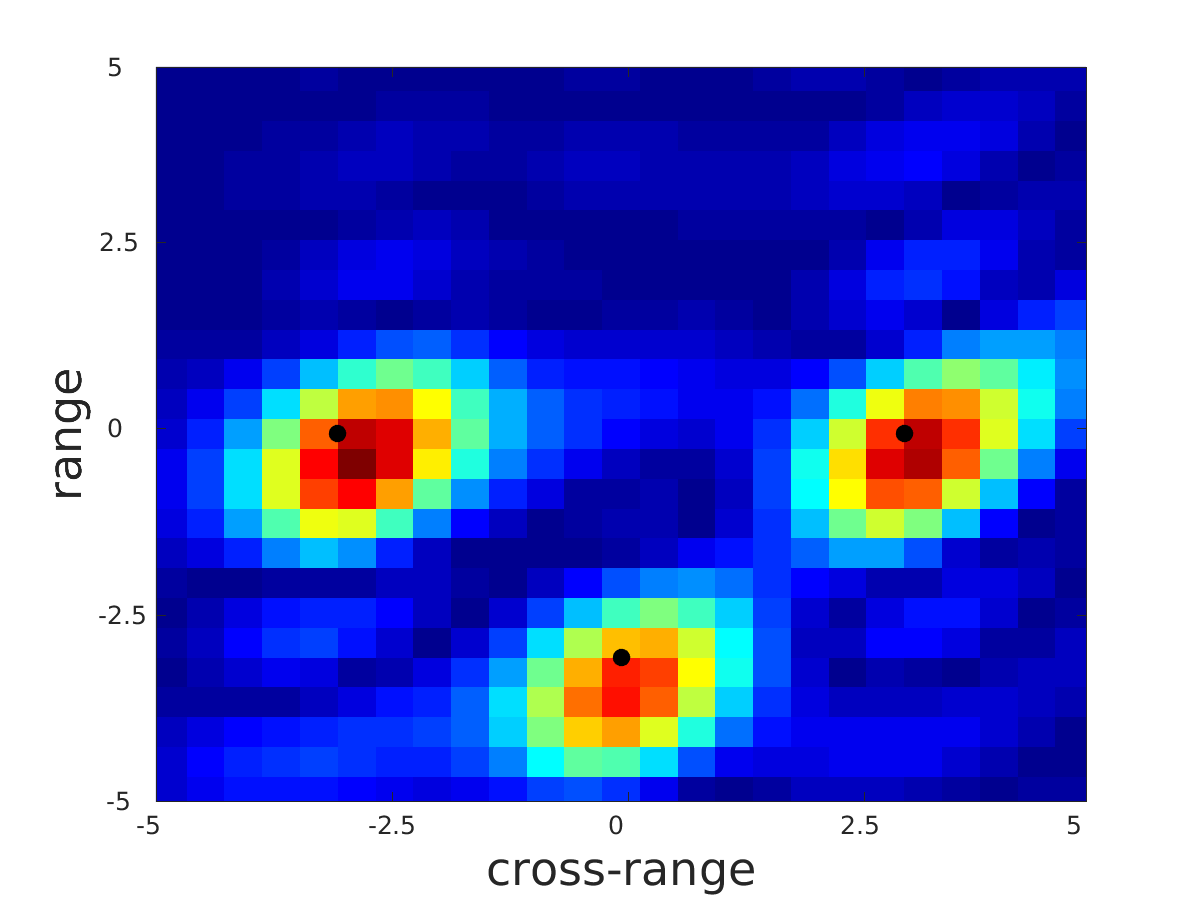}}
\caption{Passive imaging of three sources through a synthetic turbulent jet flow. Influence of the Doppler compensation factor $\smash{\Doppler}$ and refraction compensation mechanism on KM (top) and CINT (bottom) imaging functions: (a) KM image without any compensation; (b) KM image with Doppler compensation factor $\smash{\Doppler}$; (c) KM image with Doppler compensation factor $\smash{\Doppler}$ and refraction compensation mechanism. (d) CINT image without any compensation; (e) CINT image with Doppler compensation factor $\smash{\Doppler}$; (f) CINT image with Doppler compensation factor $\smash{\Doppler}$ and refraction compensation mechanism. $\smash{\FWd} = 0.3  B$ and $\smash{\SWd}=0.5  a$ for all images. The source locations are shown by dots $\bullet$. Range and cross-range dimensions are in units of $\wavelref$.}\label{fg:CINT:jet:3s}
\end{figure}

We thus aim at determining the positions of the three sources. The size of the search window is $ 10 \wavelref \times 10 \wavelref $ centered around the mid-point $\smash{(X_s, Y_s})$ of the upper sources. A pixel being of size ${2\wavelref}/{5}$, this corresponds to a window of $25\times 25$ pixels. The parameters defining the imaging setup are gathered in \tref{tb:source}. The results obtained with the KM and CINT imaging algorithms are shown in \fref{fg:CINT:jet:3s}. \alert{The theoretical range and cross-range resolutions in pixel units are $5f_\iref/2B\simeq 1$ and $5L/2a\simeq 2$ for KM, respectively, and $5f_\iref/2\FWd\simeq 4$ and $5L/2\SWd\simeq 4$ for CINT}. Again, we acknowledge that the jet flow used is ideal and does not reflect reality but we obtain the desired mechanisms with the proposed simulation. The influence of the Doppler compensation factor $\smash{\Doppler}$ is clearly seen. Without this factor the positions of the sources are significantly shifted, however using the proposed compensated imaging functions their positions are recovered with a satisfactory precision. We observe that the refraction compensation mechanism does not have any significant influence on the result. This was predictable since the speeds of sound of the two media are identical and equal to $\celm$. Furthermore the distance between the sources and the receivers is small, which leads to weak refraction of the waves. However in cases where strong refraction is likely to happen it is expected that this mechanism will become of prime importance.

\begin{table}
\begin{center}
\begin{tabular}{|c|c|c|c|c|c|}
\hline
$f_\iref$ & $X_s$ & $Y_s$ & $d$ & $L$ & $a$ \\ \hline
$50$ kHz & $12\wavelref$ & $0$ & $3\wavelref$ & $20\wavelref$ & $22\wavelref$ \\ \hline
\end{tabular}
\caption{Imaging parameters for localization of three sources through a synthetic turbulent jet.}\label{tb:source}
\end{center}
\end{table}

\section{Summary and conclusions}\label{sec:CL}

In this paper we consider array imaging techniques for moving cluttered media extending the methods that have already been developed for quiescent cluttered media in \citep{BOR05,
BOR06b}. Our aim is to apply them to classical experiments in jet flow aero-acoustics in order to localize sources through a turbulent jet, as applicable to aircraft noise characterization for example. 

After introducing a model problem and recalling the principles of two imaging algorithms for quiescent media, namely Kirchhoff migration (KM) and Coherent Interferometry (CINT), we propose in \sref{sec:func} a suitable form of these functions for moving media by introducing a Doppler compensation factor for the travel times of the array data. Accounting for this compensation it is possible to perform imaging through a heterogeneous (random) flow by simulating the back-propagation of the recorded signals in a fictitious moving ambient medium. This has the effect to partially cancel out the random time (or phase in the Fourier domain) shifts that the signals have undergone when passing through the random flow. 

In \sref{sec:Num:res} array imaging in quiescent homogeneous and random media is first introduced in order to validate the numerical tools used to generate the array data by comparisons with the existing results in \citep{BOR05}. Active imaging configurations are studied in order to localize reflectors by the KM and CINT imaging functions.
The strength of CINT images lies in the fact that they are statistically stable, which is not the case of KM images. The former are however blurred compared to the latter as a result of the smoothing that must be applied in order to obtain clear images. The trade-off between stabilization and blurring for CINT is driven by the window parameters $\SWd$ and $\FWd$ which select the cross correlations used in the imaging function. Their optimal values depend on the random medium and are therefore unknown \emph{a priori}. They are found here incrementally by successive tests, though an adaptive method to determine them optimally is developed in \citep{BOR06b}. We also note that CINT works well if significant coherence remains in the data, so that the clutter needs not be too strong. In other words, the range $L$ must be comparable to or smaller than the transport mean free path that characterizes the onset of wave diffusion, a regime where interferometric imaging cannot work. We have observed in our numerical experiments that perturbations $\sigma$ of up to about $7$ to $10$\% could be handled by the CINT algorithm. The latter is however much more demanding computationally than the KM algorithm.

In \sref{sec:moving-random-medium} the KM and CINT imaging algorithms are implemented and tested in the active imaging configuration for a heterogeneous medium with random speed of sound and moving with either a uniform or a randomly perturbed velocity. The relevance of the Doppler compensation factor is highlighted. Such correction notwithstanding, the KM imaging function is put at fault in both situations (uniform and random velocity). The three reflectors considered in these numerical experiments are however correctly localized by the CINT imaging function for a suitable choice of the window parameters $\SWd$ and $\FWd$. 

In \sref{sec:Jet} numerical tests of the KM and CINT imaging algorithms are performed to localize sources through a synthetic turbulent jet flow computed by CFD. Doppler compensation and refraction compensation at the interface between the jet layers, are implemented. In these experiments both the KM and CINT imaging functions yield accurate images of the sources. Also the refraction compensation mechanism has only a marginal influence compared to the Doppler compensation factor, which remains essential for the proper reconstruction of the images. These observations are explained by the jet configuration considered here. The different layers have the same average speed of sound, and the clutter induced by the sheared layers is mild. In other configurations with fully developed turbulent layers and heterogeneous speeds of sound refraction effects may be of prime importance, and KM imaging may be less effective than CINT imaging. Such jet configurations and other applications in moving random media should be considered in future works.

\bibliographystyle{elsarticle-num}



\end{document}